\documentclass{article}

\usepackage[authoryear]{natbib}
\usepackage[shortlabels]{enumitem}
\usepackage{hyperref}
\usepackage{amssymb}
\usepackage{float}
\usepackage{graphicx} 
\usepackage[multidot]{grffile}
\usepackage{authblk}
\usepackage[margin=0.8in]{geometry}
\usepackage{indentfirst}
\usepackage{chemformula} 

\DeclareSymbolFont{matha}{OML}{txmi}{m}{it}
\DeclareMathSymbol{\varv}{\mathord}{matha}{118}

\title{\textbf{Using HITRAN to Model Opacities for Planetary Atmospheres: Test case of Microwave Spectra of \ch{NH3}, \ch{SO2} and \ch{PH3}}}

\author[1]{Frances M. Skinner}
\author[1]{Robert J. Hargreaves}
\author[1,*]{Iouli E. Gordon}
\affil[1]{Atomic and Molecular Physics, Center for Astrophysics $\vert$ Harvard \& Smithsonian, Cambridge, MA USA}
\affil[*]{Corresponding author E-mail address: igordon@cfa.harvard.edu}

\date{\vspace{-4mm}\small\today}

\begin{document}
\maketitle
\vspace{-6mm}

\renewenvironment{abstract}
 {\small
  \begin{center}
  \bfseries \abstractname\vspace{-.5em}\vspace{0pt}
  \end{center}
  \list{}{%
    \setlength{\leftmargin}{0.5mm}
    \setlength{\rightmargin}{\leftmargin}
}
  \item\relax}
 {\endlist}
 \vspace{4mm}
\begin{abstract}
\centering
The latest version of the HITRAN molecular spectroscopic database, HITRAN2020, has recently been released featuring many updates, including line-by-line broadening parameters (and their temperature dependence) appropriate for the dominant constituents of planetary atmospheres.  In this work, line shape codes suitable for calculating microwave spectra have been implemented within the HITRAN Application Programming Interface (HAPI). These new additions allow for spectroscopic calculations of microwave absorbing species pertinent to current and future studies of the atmospheres of Jupiter and Venus, and more generally for the atmospheres of gas giants and rocky planets. The inversion spectrum of the \ch{NH3} molecule broadened by \ch{H2}, \ch{He} and \ch{H2O} dominates the microwave region of Jupiter. Whereas for Venus, accurate spectroscopic data of \ch{SO2} broadened by \ch{CO2} is necessary in order to determine its significance, if any, on the reported detection of \ch{PH3} in the Venusian upper cloud deck. Comparisons have been made to available microwave laboratory opacities and the following results illustrate that HITRAN data can be used in conjunction with HAPI to reproduce the existing experimental measurements and provide reliable calculation of planetary opacities. Users should be mindful regarding selection of appropriate parameters in HITRAN and selecting suitable line shape functions in HAPI, depending on the spectral region, target molecular species, as well as ambient chemical and thermodynamic conditions. 
\end{abstract}

\section*{\small Introduction}\vspace{-1.5mm}\hrule\vspace{3mm}\label{sec:spectroscopy} 
The high-resolution transmission molecular spectroscopic database (HITRAN) is an international standard for reference molecular spectroscopy in various fields of science and industry, where its primary goal is to interpret absorption of electromagnetic radiation in terrestrial, planetary and exoplanetary atmospheres. HITRAN2020 \citep{10.1016/j.jqsrt.2021.107949}, the most recent edition, contains a compilation of line-by-line spectroscopic parameters that can be used to simulate and analyze the transmission and emission spectra of 55 molecules in gaseous media \citep{10.1016/j.jqsrt.2021.107949}. Furthermore, HITRAN contains experimental absorption cross sections for over 300 additional molecules \citep{10.1016/j.jqsrt.2019.04.001}, as well as collision induced absorption \citep{10.1016/j.icarus.2019.02.034} for many important collisional complexes. The HITEMP database follows the same formalism as HITRAN but contains substantially more transitions that are necessary for high-temperature calculations, including those encountered in the Jovian and Venusian atmospheres. The HITRAN Application Programming Interface (HAPI) \citep{10.1016/j.jqsrt.2016.03.005} contains a set of Python scripts that have been developed for use with HITRAN data, thus allowing line-by-line calculations of absorption cross sections, transmission and emission spectra, and more. 

Prior to the work of \citet{10.1016/j.jqsrt.2015.09.003}, HITRAN provided only air-broadening (typically from a combination of \ch{N2}- and \ch{O2}-broadening) and self-broadening parameters for every spectral line. However, line broadening is strongly dependent on the perturber and if one aims at accurate modeling of the spectra of atmospheres dominated by gases other than \ch{N2} and \ch{O2}, the air-broadening values cannot be employed. For example, \citet{10.1029/2019JD030929} demonstrated the significant differences in absorption cross-sections calculated using air-broadening and \ch{H2O}-broadening parameters. Since the atmospheres of gas giants share \ch{H2}, \ch{He}, and (to a lesser extent) \ch{H2O} as major constituents, and since \ch{CO2} predominates in atmospheres of Venus and Mars (and exoplanets with a history of high-volcanic activity), efforts have been made towards incorporating broadening parameters due to collisions with these gases into HITRAN for planetary atmosphere applications \citep{10.1016/j.jqsrt.2015.09.003, 10.1029/2019JD030929,yansplanetarybroadeners}. These studies combine numerous experimental and theoretical data in order to determine semi-empirical models of collisional broadening for each molecule. These models thereby allow effective application of the relevant broadening parameters to each spectroscopic transition of the molecules studied \citep{yansplanetarybroadeners}. 

It is worth defining here that "broadening parameters" refer to 1) Lorentzian half-widths at half maximum (HWHM) denoted as $\gamma_{\ch{H2}}$, $\gamma_{\ch{He}}$, $\gamma_{\ch{CO2}}$, and $\gamma_{\ch{H2O}}$ for \ch{H2}-, \ch{He}-, \ch{CO2}-, and \ch{H2O}-broadening, respectively; 2) temperature dependence (exponent) of these half-widths (denoted as $\eta_{\ch{H2}}$, $\eta_{\ch{He}}$, $\eta_{\ch{CO2}}$, $\eta_{\ch{H2O}}$) defined through the power law given in equation~\ref{eq:pl}; 3) collisional line shifts (denoted as $\delta_{\ch{H2}}$, $\delta_{\ch{He}}$, $\delta_{\ch{CO2}}$, $\delta_{\ch{H2O}}$) which at the moment are available only for some HITRAN molecules and in some cases only for some specifically selected lines of these molecules. 
The power law equation for determining the HWHM at T is given as:
\begin{equation}
\centering
\resizebox{.2\hsize}{!}{$\gamma(T)=\gamma(T_0)\left[\frac{T_0}{T}\right]^\eta$}
\label{eq:pl}
\end{equation}
where T$_0$ is the reference temperature (296~K in HITRAN) and $\gamma (T_0)$ is the HWHM at the reference temperature. 

where T$_0$ is the reference temperature (296K in HITRAN) and $\gamma (T_0)$ is the HWHM at the reference temperature. 

HAPI is capable of utilizing the aforementioned parameters to calculate absorption cross sections of gaseous spectra with a variety of gas mixture concentrations. This work demonstrates how HITRAN2020 data can be used in conjunction with HAPI to reproduce available experimental absorption data, obtained under Jovian and Venusian conditions. The ability to reproduce experimental opacities in turn provides confidence in the quality and applicability of the calculations using HITRAN data under planetary conditions. The primary addition applied to HAPI in order to carry out this work, was to include additional asymmetric line shape functions appropriate for modeling in the microwave spectral region. These new (to HAPI) line shape functions are available as a Python script to be used in conjunction with HAPI and are given in the supplementary materials section in conjunction with publication of this work in the Monthly Notices of the Royal Astronomical Society (MNRAS).

\begin{center}
\subsection*{\normalsize Choice of Test Cases}
\end{center}

\subsubsection*{\small Inversion Spectrum of \ch{NH3} and the Jovian Atmosphere}\vspace{-1.5mm}\hrule\vspace{3mm}\label{sec:jovian} 
Missions to Jupiter equipped with spectrometers are necessary in order to better understand the atmospheric composition of the gas giant. The microwave radiometers are of especial advantage because Jupiter's neutral atmosphere is shrouded by clouds that are impenetrable to all but microwave radiation \citep{2018EPSC...12..204O}. Ground-based microwave measurements of Jupiter are impeded by intense synchrotron radiation that obscures all except the shortest-wavelength microwave radiation emitting from above the few-bar pressure level of the atmosphere \citep{2018EPSC...12..204O}. In August of 2011 NASA launched its ``New Frontiers" Juno mission with a primary objective to better understand the complexities of Jupiter's atmosphere \citep{10.1002/2017GL074277}. Juno is spin-stabilized and solar-powered \citep{2019AGUFM.P44A..03L}, and carries as one of its payloads a microwave radiometer (MWR) that can be used to retrieve the atmospheric composition in the Jovian atmosphere \citep{10.1002/2017GL074277}. The Juno spacecraft is passing below Jupiter's radiation belts \citep{10.1126/science.abf1015}. In addition, the MWR has the capability to separate Jupiter's synchrotron emission effectively from atmospheric thermal emission \citep{10.1029/2021JE006858}.

Juno's MWR operates within six microwave channels to retrieve the abundance of molecular constituents from the microwave signature of Jupiter \citep{10.1007/s11214-017-0349-5,10.1126/science.aal2108}. At these frequencies, the microwave opacity of the Jovian atmosphere is dominated by the inversion of the \ch{NH3} molecule and is broadened by the major species in Jupiter's atmosphere \ch{H2} and \ch{He} as well as the minor species \ch{H2O} \citep{10.1016/S0032-0633(02)00144-7}. Therefore, it is essential to correctly model the \ch{NH3} microwave spectrum in order to maximize the benefit of Juno's observations. 

Recently, data from Juno was used to determine that the mid-latitudes of Jupiter's atmosphere show a significant depletion of \ch{NH3} \citep{2019EPSC...13.1142G}, while Jupiter's equatorial zone displays a column of enriched \ch{NH3} gas \citep{10.1002/2017GL073159,10.1002/2017GL074277}. It was suggested by \citet{10.1029/2020JE006403} that this strong \ch{NH3} vertical distribution is due to a lack of ``mushballs'' (i.e., mixed-phase \ch{NH3}/\ch{H2O} condensates \citep{10.1016/0019-1035(73)90019-5}) and storms in the equatorial zone \citep{10.1029/2021JE006858}. Although, with the Juno mission underway, the detailed composition of Jupiter's atmosphere is still under investigation. The MWR is able to shed light on the mysteries of Jupiter's atmosphere since it can probe the composition deeper than what is possible from any Earth-based telescope or previous spacecraft \citep{2019AGUFM.P21G3455B}. This has enabled better understanding of the mixture and concentration of \ch{H2O} and \ch{NH3} in Jupiter's atmosphere and in particular has furthered understanding of the \ch{NH3} plume in Jupiter's equatorial zone. Although the global distributions of winds, heat ﬂuxes, temperature gradients, and chemical species below the weather layer are still largely unknown \citep{10.1002/2017GL074277} they remain under investigation by the Juno mission.

For this study, \ch{NH3} absorption under Jovian conditions is considered with \ch{He}, \ch{H2} and \ch{H2O} broadening applied. Overall, there are a variety of molecular species in HITRAN which have broadening parameters relevant for the Jovian atmosphere. For instance, HITRAN provides \ch{He}, and \ch{H2}-broadening and their temperature dependence exponents for \ch{HCN} which was observed by Cassini/CIRS in December 2000 on Jupiter \citep{10.1016/j.icarus.2006.05.018}. Additional molecular species which have been detected on Jupiter are detailed in Table~1 of \citet{10.1016/S0032-0633(02)00144-7}. These include \ch{H2O}, \ch{CO2}, \ch{CO}, \ch{CH4}, \ch{HCl}, \ch{C2H2}, \ch{PH3}, \ch{H2S}, \ch{H2} and \ch{GeH4}. In HITRAN these molecules are available with some or all \ch{H2}, \ch{He}, \ch{H2O} and \ch{CO2} broadening parameters, temperature dependence exponents and for select molecules there are pressure-induced line shift values. Molecules that have also been detected on Jupiter and detailed in \citet{10.1016/S0032-0633(02)00144-7,10.1029/95GL01718} which are available in HITRAN but without ``planetary" broadening parameters include; \ch{C2H6}, \ch{C2H4}, \ch{C4H2} and \ch{CS2}. 

A comprehensive set of laboratory studies at the Georgia Institute of Technology have investigated the microwave opacity of \ch{NH3} under Jovian conditions \citep{10.1016/0019-1035(87)90116-3, 10.1016/0019-1035(89)90059-6, 10.1029/91JE01740, 10.1016/j.icarus.2003.09.003, 10.1029/2003JE002201, 10.1016/j.icarus.2009.02.002, 10.1016/j.icarus.2010.12.010, 10.1016/j.icarus.2014.06.017, 10.1016/j.icarus.2016.07.013}. These results have been fundamental in the development of \ch{NH3} opacity models and have contributed to the interpretation of Juno's atmospheric observations \citep{10.1029/2021JE006858}. These laboratory measurements have formed the basis of a series of \ch{NH3} opacity models following the method of \citet{10.1016/j.icarus.2009.02.002}, with the most recent modifications presented in \citet{10.1016/j.icarus.2016.07.013}. To clarify, it is not within the scope of this work to compare to Radiative Transfer Models which are utilized by the Juno mission. Instead, only the \ch{NH3} opacity models and their corresponding laboratory measurements will be used in comparison with HITRAN2020 data and HAPI. 

\subsubsection*{\small Microwave Spectra of \ch{SO2} \& \ch{PH3} and the Venusian Atmosphere}\vspace{-1.5mm}\hrule\vspace{3mm}\label{sec:venusian} 
\ch{SO2} is a principal constituent in the sulfur cycle on Venus \citep{10.1016/j.icarus.2017.05.003}. Its spectrum is broadened by \ch{CO2}, which makes up the majority of the Venusian atmosphere. It is thought that, due to Hadley cell circulation and convective transportation, \ch{SO2} effectively transfers from the lower to the upper portions of the Venusian atmosphere \citep{10.1016/j.icarus.2017.05.003}. In the middle/upper atmosphere, \ch{SO2} is oxidized and forms \ch{SO3} which then synthesizes with \ch{H2O} to precipitate as \ch{H2SO4} droplets that enrich the Venusian clouds and haze \citep{10.1016/j.icarus.2017.05.003}. This process leads to a substantial decrease in \ch{SO2} abundance above the Venusian clouds \citep{10.1016/j.icarus.2017.05.003}. The main cloud deck is placed around altitudes of 45-70~km \citep{10.1093/acrefore/9780190647926.013.4} with \ch{SO2} as the most abundant sulfur-bearing species in the middle \citep{10.1029/176GM06} and lower atmosphere, varying with altitude, latitude and time \citep{10.1051/0004-6361/202140837}.

\ch{PH3} has tentatively been observed by \citet{10.1038/s41550-020-1174-4,10.1038/s41550-021-01423-y} within the Venusian atmosphere. This potential detection was based on two separate observations using the James Clerk Maxwell Telescope (JCMT) and the Atacama Large Millimeter/submillimeter Array (ALMA). \citet{10.1038/s41550-020-1174-4} show that the observed feature can be attributed to the \ch{PH3} rotational transition for the (\emph{$J'_{K'}$}$\leftarrow$\emph{$J''_{K''}$}) 1$_{0}$$\leftarrow$$0_{0}$ line located at 8.904~cm$^{-1}$ (\emph{J} is the rotational quantum number associated with the total angular momentum and \emph{K} is associated with the projection of the total angular momentum on the symmetry axis of the molecule), originating from the middle/upper cloud deck layers with altitudes approximately 53--61~km. On the one hand, a detection of \ch{PH3} is not unusual elsewhere in the Solar System; for instance, \ch{PH3} is also known to be present in Jupiter’s atmosphere \citep{10.1016/j.icarus.2009.03.023, 10.1086/155009,10.1086/154570}. On the other hand, \ch{PH3} is an unexpected species in the Venusian atmosphere since there are no known photochemical pathways where \ch{PH3} can be abiotically produced with the proposed abundances \citep{10.1038/s41550-020-1174-4, 2020arXiv200906499B}. The primary significance of \ch{PH3} on Venus is that it has been proposed as a potential biosignature by \citet{10.1089/ast.2018.1954}. Furthermore, Venus has a highly oxidized atmosphere, therefore the presence of \ch{PH3} is unexpected since it is not an oxidized species \citep{10.1051/0004-6361/202039559,2020arXiv200906499B}.

However, it has been proposed that the spectral feature, attributed to \ch{PH3} by \citet{10.1038/s41550-020-1174-4}, could be explained by the rotational (\emph{$J'_{K_{a}',K_{c}'}$}$\leftarrow$\emph{$J''_{K_{a}'',K_{c}''}$}) transition 30$_{9,21}\leftarrow$ 31$_{8,24}$ of \ch{SO2}, as both molecules have absorption lines that coincide with the identified absorption feature at 8.904~cm$^{-1}$ or 266.94~GHz \citep{10.3847/2041-8213/abde47}. Here \emph{J, K$_{a}$, K$_{c}$} are rotational quantum numbers associated with asymmetric top molecules. In the microwave region the \ch{PH3} spectra demonstrates little opacity before the 1$_{0}\leftarrow$ 0$_{0}$ line. For \ch{SO2} there are many dense spectral transitions at the same spectral region as the \ch{PH3} 1$_{0}\leftarrow$ 0$_{0}$ line. 

For these reasons, this work concentrates on \ch{SO2} and \ch{PH3} under Venusian conditions. With that, as it will be widely discussed below, HITRAN contains \ch{CO2}-broadening parameters for the lines of sulfur dioxide, while they are not available for phosphine due to the lack of measurements in calculations. It should be noted that, in general, there are a number of chemical species which have been detected on Venus (see, for instance, \citet{10.1029/2020GL091327}) available in HITRAN. These HITRAN data and their available perturbers can be used to generate planetary opacities under a variety of temperatures, pressures, and mixture concentrations. There are some broadening considerations to keep in mind, though. 

Many of these molecules have broadening parameters relevant to the Venusian atmosphere, but some do not. For example, \ch{HF}, \ch{HCl} and \ch{OCS} have \ch{CO2}-broadening parameters (as well as \ch{H2}, and \ch{He}). Naturally, HITRAN features a very comprehensive line list for 12 isotopologues of carbon dioxide itself, and one can obtain \ch{CO2}-broadening parameters by selecting self-broadening. For self-broadening of \ch{CO2}, HITRAN contains some additional data in comparison with other molecules. This includes line-mixing parameters and their temperature dependencies as well as non-Voigt parameters (\citet{ 10.1016/j.jqsrt.2020.107283}). In the context of this paper, this may not have much of an impact as since the symmetric isotopologues of \ch{CO2} there is no permanent dipole moment; therefore, pure-rotational transitions in the microwave (MW) region are forbidden, and only very small dipole moment of minor asymmetric isotopologues allows some very weak transitions for these species.

Molecules that have also been detected in the Venusian atmosphere which are available in HITRAN but without \ch{CO2}-broadening parameters include; \ch{H2O}, \ch{O3}, \ch{O2}, \ch{OH} , \ch{ClO}, \ch{N2}, \ch{O}, \ch{SO3} and \ch{SO} \citep{Krasnopolsky2013, 10.1051/0004-6361:200809761,10.1029/2020JE006625, 10.1029/2020GL091327,10.1016/j.icarus.2018.04.022,10.1029/2020GL091327,10.1016/j.icarus.2012.09.007,10.1016/j.icarus.2010.02.013}. Some of these molecules have broadening parameters due to pressure of water vapor, hydrogen and helium, but these are not relevant to Venus. There is an extensive effort underway to provide \ch{CO2}-broadening parameters for the lines of water vapor. However, for other molecules there is a lack of measurements and calculations and we strongly advocate for relevant experimental and theoretical works. 

This work uses HITRAN2020 data (which contains state-of-the-art spectral parameters, including broadening parameters due to pressure of different gases) and line shape codes introduced into HAPI to calculate \ch{SO2} broadened by \ch{CO2} under the conditions at Venus. Comparisons of this work against laboratory opacities by \citet{10.1016/j.icarus.2015.03.028,10.1016/j.icarus.2014.09.012} are presented. The available microwave laboratory measurements of the opacity of \ch{SO2} in a \ch{CO2} atmosphere were conducted under simulated conditions of the Venusian atmosphere. These laboratory measurements mainly cover the very low frequency rotational transitions for \ch{SO2}. Regrettably, these measurements do not coincide with the spectral region where \ch{PH3} was detected on Venus. Nevertheless, this works analyses can validate general applicability of new HITRAN data and the microwave line shapes implemented into HAPI in the course of this work.

The potential detection of \ch{PH3} on Venus has resulted in a renewed interest in Venus with forthcoming missions focusing on observing the Venusian atmosphere. Notably, one of these upcoming missions is the orbiter and atmospheric balloon mission Shukrayaan-1 by the Indian Space Research Organization (ISRO) \citep{2018cosp...42E1349H} which will observe the surface and atmospheric composition of Venus in 2024. Additionally, the orbiter and lander Venera-D by NASA-Space Research Institute (IKI)/Roscosmos \citep{2021cosp...43E.251Z} will be exploring Venus when it launches in 2029. The DAVINCI+ (Deep Atmosphere Venus Investigation of Noble gases, Chemistry, and Imaging) NASA mission will be analyzing the atmospheric chemistry of Venus in 2029 using near-IR, IR and ultra-violet (UV) measurements \citep{2021cosp...43E.442G}. Due to the numerous upcoming missions to observe and analyze the Venusian atmosphere, it is necessary that HITRAN provide more accurate spectroscopic data of relevance to Venus.

Since the primary chemical constituent in the Venusian atmosphere is \ch{CO2} \citep{10.1093/acrefore/9780190647926.013.4}, the ideal situation would have been for this work to compare against laboratory opacities of \ch{PH3} broadened by \ch{CO2}. However, no such measurements exist, and there are no relevant data in HITRAN either, due to lack of measurements or calculations. However, there are data for \ch{H2} and \ch{He} broadening of \ch{PH3} lines in HITRAN. Also there are experimental measurements of microwave opacities of \ch{PH3} in a \ch{H2} and \ch{He} atmosphere conducted by \citet{10.1029/2003JE002201,10.1016/j.icarus.2003.09.003,10.1006/icar.1999.6158,10.1006/icar.2001.6622}. Therefore, here to test the general approach we compare these experimental opacities in a \ch{H2} and \ch{He} atmosphere with the ones calculated using the HITRAN data. In the meantime, we encourage experimental and theoretical studies of \ch{PH3} lines perturbed by \ch{CO2}.

\section*{\small Methods}\vspace{-1.5mm}\hrule\vspace{3mm}\label{sec:linelists} 
For accurate line-by-line opacity calculations, the line positions, intensities, line broadening parameters and line shape details are required for each constituent molecule. The following summarizes the microwave data available in HITRAN2020 for the selected molecules included in this work (i.e., \ch{NH3}, \ch{SO2} and \ch{PH3}). Many updates (with respect to previous editions of the database) which are relevant for calculating opacities other than \ch{NH3}, \ch{SO2} and \ch{PH3} for planetary atmospheres are detailed in the HITRAN2020 paper \citep{10.1016/j.jqsrt.2021.107949}. It is important to note that the HITRAN line intensities account for ``natural'' terrestrial isotopic abundances \citep{10.1016/j.jqsrt.2021.107949,10.1063/1.555720}. Thus, the calculations presented in this work use these natural abundance intensities. However, when calculating opacities for Jovian or Venusian observations one will need to account for the abundances characteristic to those environments (see for instance \citet{10.1023/A:1024633603624}). 

The HITRAN2020 paper details improvements (with respect to previous editions of the HITRAN database) to the \ch{SO2} line list \citep{10.1016/j.jqsrt.2021.107949} relevant to the microwave region. For instance, the principal isotopologue \ch{^{32}S^{16}O2} has newly included microwave hot bands which came from the work of \citet{so2_naumenko_2019} and greatly increased the number of lines in this region. Moreover, the minor isotopologue \ch{^{34}S^{16}O2} line list contains microwave hot band additions, which came from the work of \citet{10.1016/j.jqsrt.2018.11.039}. Furthermore, two minor isotopologues \ch{^{33}S^{16}O2} and \ch{^{16}O^{32}S^{18}O} have also been added to HITRAN from \citet{10.1016/j.jqsrt.2018.11.039}. 

The \ch{NH3} line list within the microwave region in HITRAN contains the inversion transitions, pure rotational transitions, their hot bands, as well as analogous transitions of the minor isotopologue \ch{^{15}NH3}. The inversion mode of \ch{^{14}NH3} dominates the microwave region with a maximum at around 0.8~cm$^{-1}$ (24 GHz). Similarly, transitions between the two inverted forms of the $\nu_{2}$ vibration are centered around 35~cm$^{-1}$, in HITRAN, these transitions are provided by \citet{pearsonnh3}. Additional pure rotational lines and a number of hot bands are included in HITRAN from the reanalysis of \ch{NH3} by \citet{10.1016/j.jqsrt.2013.05.027}. Moreover, in HITRAN2020 selected lines belonging to \ch{^{14}NH3} have been added from the ExoMol CoYuTe line list \citep{10.1093/mnras/stz2778} after the MARVEL process was applied to energy levels (described in \citet{10.1016/j.jqsrt.2020.107027}). The inversion band for the \ch{^{15}NH3} molecule is also included in HITRAN based on the CDMS database \citep{10.1016/j.molstruc.2005.01.027, 10.1016/j.jms.2016.03.005} and (taking isotopic abundance into account) is approximately three orders of magnitude weaker than the principal isotopologue near 0.8~cm$^{-1}$. Additional, \ch{^{15}NH3} line position data comes from the work of \citet{10.1063/1.5088751} who investigate the $\nu_{2}$ = 1, 2 and $\nu_{4}$ = 1 bending states of \ch{^{15}NH3}. Also, some intensity data of \ch{^{15}NH3} comes from the work of \citet{10.1016/j.jqsrt.2014.10.023} who generated a theoretical room-temperature line list. 

Finally, with regards to \ch{PH3}, the data available in the microwave range primarily originates from the CDMS database \citep{10.1016/j.molstruc.2005.01.027,10.1016/j.jms.2016.03.005} and includes the weak inversion spectral lines and stronger rotational lines of \ch{PH3}.

\subsection*{\small Relevant Broadening Parameters in HITRAN}\vspace{-1.5mm}\hrule\vspace{3mm}\label{sec:broadeners} 
Historically, HITRAN contains ``traditional'' air- and self-broadening parameters for all transitions of each molecule, however in the last two editions the broadening by gases dominating planetary atmospheres have been introduced for many molecules. HITRAN data are commonly used by the planetary community, for example the Planetary Spectrum Generator (PSG) \citep{10.1016/j.jqsrt.2018.05.023} enables users to calculate opacities using the available broadening species from HITRAN. This sections details the additional broadening parameters, which are applicable for planetary atmospheres, for \ch{SO2}, \ch{NH3}, and \ch{PH3}. Here, line broadening by \ch{X} is denoted $\gamma_{\ch{X}}$, with a temperature dependence $n_{\ch{X}}$, and a line shift denoted $\delta_{\ch{X}}$ (e.g., $\gamma_{\ch{H2}}$, $n_{\ch{H2}}$, and $\delta_{\ch{H2}}$ refer to the \ch{H2}-broadening parameters).

Additional broadening parameters for \ch{NH3} are provided in HITRAN for \ch{H2}, \ch{He}, \ch{CO2} \citep{10.1016/j.jqsrt.2015.09.003} and \ch{H2O} \citep{10.1029/2019JD030929}. HITRAN collates broadening data through a detailed process of validation and studies; the work done in \citet{10.1016/j.jqsrt.2015.09.003} gave rise to the inclusion of \ch{H2}, \ch{He} and \ch{CO2} broadening of \ch{NH3} in HITRAN. The \ch{NH3} broadened by \ch{CO2} data could potentially be relevant for Venusian conditions due to the recent tentative identification of \ch{NH3} on Venus by \citet{10.1029/2020GL091327} from the re-examination of the Pioneer Venus Large Probe Neutral Mass Spectrometer. As discussed by \citet{10.1016/j.jqsrt.2015.09.003}, the \ch{NH3}-\ch{H2} and \ch{NH3}-\ch{He} broadening parameters in different vibrational bands were studied quite extensively in several peer-reviewed publications. Overall, the broadening coefficients by \ch{H2} and He exhibited a strong dependence on the rotational quantum numbers ($J$ and $K$). Each of the broadening parameters in HITRAN have been fitted to a rotationally-dependent empirical function proposed by \citet{10.1016/S0022-4073(02)00354-0}. The data in HITRAN for shifting of \ch{NH3} lines by \ch{He} were obtained from peer-reviewed works and incorporated into HITRAN after detailed analysis and verification. The shifts are known to exhibit both rotational and vibrational dependence. The data in HITRAN for shifting of \ch{NH3} lines by \ch{H2} and \ch{He} were obtained from peer-reviewed works and incorporated into HITRAN after detailed analysis and verification. Limited extrapolations were applied, keeping in mind strong vibrational dependence.

\ch{H2O} vapor is one of the most effective broadener's when compared to the other species in this work. It was noted by \citet{10.1007/s11214-016-0265-0} that \ch{H2O} vapor broadens the absorption spectrum of \ch{NH3} about five times more than the equivalent amount (by volume) of molecular \ch{H2}, and about nine times as much as \ch{He}. In the work by \citet{10.1029/2019JD030929}, the \ch{H2O} vapor pressure‐broadening parameters (and their temperature‐dependent exponents) were evaluated for the species \ch{CO2}, \ch{N2O}, \ch{CO}, \ch{CH4}, \ch{O2}, \ch{NH3} and \ch{H2S} from both experimental and theoretical studies. The \ch{H2O} broadening data in the literature are substantially scarcer than that for \ch{H2} and \ch{He}. Nevertheless, available measurements were used to create semi-empirical models to populate the entire line lists of the aforementioned molecules into the HITRAN database with relevant parameters \citep{10.1029/2019JD030929}. In many cases the \ch{H2}, \ch{He}, \ch{CO2} and \ch{H2O} broadening data in HITRAN is based on a combination of theoretical and experimental values, distinctions between isotopologues were made whenever the necessary data were available in the literature. Otherwise, it was assumed that non-deuterated minor isotopologues have the same foreign broadening coefficients as the principal isotopologue \citep{10.1016/j.jqsrt.2015.09.003}. Note that there have been laboratory studies on the influence of pressure-broadening from \ch{CH4} on \ch{NH3} in the microwave region. These comparisons are not conducted in this work, partly due to the lack of \ch{CH4} broadening data on \ch{NH3} inversion lines. Also, due to relatively low abundance of \ch{CH4} in the atmosphere of Jupiter (0.2\% by volume), its effect on the microwave spectrum, which is observed by the Juno MWR, will be minimal \citep{2012DPS....4441228C}. 

\citet{10.1016/j.jqsrt.2015.09.003} noted that the \ch{CO2}-broadening parameters of \ch{SO2} in HITRAN2016 were based on a rudimentary approximation due to extreme lack of data at that time. In brief, the value corresponding to one measurement of a single transition \citep{10.1063/1.1733747} was used to derive a scale factor from comparison to the $\gamma_{\ch{H2}}$ value for the same transition. Additionally, the HITRAN2016 broadening data used a constant 0.75 value for the temperature dependence of \ch{SO2} broadened by \ch{CO2} and the pressure shift was also not accounted for and instead given a constant value of zero \citep{10.1016/j.jqsrt.2015.09.003}. Nevertheless, it was found by \citet{10.3847/2041-8213/abde47} that the HITRAN2016 \ch{CO2} broadening data of \ch{SO2} were broadly consistent with laboratory results from \citet{10.1016/j.icarus.2015.03.028}. 

For HITRAN2020, the \ch{SO2}-\ch{CO2} broadening data were updated based on the experimental measurements from \citet{10.1016/J.JQSRT.2018.12.030} and \citet{10.1016/j.jqsrt.2017.02.018}. Furthermore, extrapolated (beyond laboratory measurements) broadening, temperature dependence, and line shift values were applied by \citet{10.1016/J.JQSRT.2018.08.022} to transitions where broadening data were not directly available. This process has provided unique \ch{CO2} broadening, temperature dependence and pressure induced line shift (which was previously not available in HITRAN) data to every \ch{SO2} transition. A full in-depth analysis as to the validity of the new \ch{SO2}-\ch{CO2} broadening data are available in \citep{10.1016/J.JQSRT.2018.08.022} and there it was found that the uncertainty between the applied theoretical values from experimental measurements are less than 8\%. Needless to say, the updated \ch{SO2}-\ch{CO2} broadening data from \citet{10.1016/J.JQSRT.2018.08.022} are expected to give even better performance than that of HITRAN2016. 

The \ch{H2} and \ch{He} broadening data for \ch{PH3} were added for HITRAN2020 and described in \citet{yansplanetarybroadeners}. The \ch{PH3}-\ch{H2} broadening is based on an empirical polynomial fit to experimental data and the \ch{PH3}-\ch{He} broadening data have been obtained from a linear fit to available experimental data. 

\subsection*{\small Line Shape Calculations}\vspace{-1.5mm}\hrule\vspace{3mm}\label{sec:lineshapes}
The line shape used for spectral calculations in the microwave region is extremely important. This is particularly true for \ch{NH3} because of the inversion mode, which provides a large number of densely packed, low-frequency transitions.
Symmetric line shapes, including Lorentzian and Voigt profiles fail to reproduce the opacities of microwave absorbing species at very low frequencies \citep{10.1017/9781108768825}. Consequently, as part of this work, widely-used asymmetric line shape profiles have been added to HAPI. 

\citet{10.1103/RevModPhys.17.227} used a simple derivation from the \citet{1929pomo.book.....D} and \citet{1915KNAB...18..134L} theories of line-broadening for a harmonic oscillator, with a special focus on the microwave region. The Van Vleck \& Weisskopf spectral intensity at $\nu$ (in cm$^{-1}$) can be written as
\begin{equation}
\centering
\resizebox{.5\hsize}{!}{$F(\nu,\nu_0,\gamma_0)=\left(\frac{1}{\pi}\right)\left(\frac{\nu}{\nu_0}\right)^2\left[\frac{\gamma_0}{\left(\gamma_0^2+(\nu_0-\nu)^2\right)}+\frac{\gamma_0}{\left(\gamma_0^2+(\nu_0+\nu)^2\right)}\right]$}
\label{eq:vvw}
\end{equation}
where $\nu_0$ is the unperturbed line position in cm$^{-1}$ and $\gamma_0$ is the speed-averaged line-width in cm$^{-1}$/atm.

The Van Vleck \& Weisskopf (VVW) line shape is applicable to most microwave absorbing species, but was initially created and compared to the inversion spectrum of \ch{NH3} to act as a more compatible line shape, since the asymmetry allows it to better model the complicated spectrum. Following this study, \citet{10.1038/157478a0} reached the same conclusion as \citet{10.1103/RevModPhys.17.227}, that the well-known Lorentz formula for the shape of a spectral line is not sufficient for the microwave spectra of \ch{NH3}, except at frequencies close to the resonance frequency. Thus, \citet{10.1038/157478a0} adapted the VVW theory to account for harmonic oscillators. Additionally, \citet{10.1103/PhysRev.76.647} created a generalized theory of collision broadening to predict line widths in the microwave and IR regions, their results showed successful comparisons with experiments of self-broadened \ch{NH3} in the microwave region. It was not until \citet{10.1063/1.1698661} that it was determined that the VVW line shape inadequately accounted for low-frequency wings of the \ch{NH3} inversion lines. In particular, \citet{10.1063/1.1698661} found the VVW line shape is $\sim$40\% too weak at 0.094~cm$^{-1}$ and is also not comparable at 0.31~cm$^{-1}$ when evaluated alongside experimental measurements at pressures of 0.06-0.39~atm and temperatures from 261-373~K.

An alternative asymmetric line shape in the microwave region is the formalism derived by \citet{10.1103/PhysRev.97.395}. The Gross line shape formula resulted in a model for a friction-damped oscillator and is written as
\begin{equation}
\centering
\resizebox{.4\hsize}{!}{$F(\nu,\nu_0,\gamma_0)=\left(\frac{1}{\pi}\right)\left(\frac{\nu}{\nu_0}\right)\left[\frac{4\nu\nu_0\gamma_0}{\left(4\nu^2\gamma_0^2+(\nu_0^2-\nu^2)^2\right)}\right]$}
\label{eq:gross}
\end{equation}
where $\nu_0$ and $\gamma_0$ are the same parameters as described in equation~(\ref{eq:vvw}). The polarization hypothesized by \citet{10.1103/PhysRev.97.395} is reached as a result of kinematic motion between collisions. Although, both the VVW and Gross line shapes converge at the line center, the Gross line shape sheds light on the processes responsible for the shapes of the rotational spectral lines in gases in the microwave region. 

Finally, \citet{10.1103/PhysRev.145.7} developed a line shape calculation for reproducing the unique shape of the strong inversion spectra of \ch{NH3}. \citet{10.1103/PhysRev.145.7} derived a new formula for the shape of resonance lines, as a function of frequency and pressure, which takes into account the special effect of collision-induced transitions between the upper and the lower levels of the line. This Ben-Reuven line shape is given as
\begin{equation}
\centering
\resizebox{.6\hsize}{!}{$F(\nu,\nu_0,\gamma_0,\zeta_0,\delta_0)=\left(\frac{2}{\pi}\right)\left(\frac{\nu}{\nu_0}\right)^2\left[\frac{\nu^2(\gamma_0-\zeta_0)+(\gamma_0+\zeta_0)\left[\gamma_0^2-\zeta_0^2+(\delta_0+\nu_0)^2\right]}{4\gamma_0^2\nu^2+\left[\nu^2-{\gamma_0^2}+\zeta_0^2-(\delta_0+\nu_0)^2\right]^2}\right]$}
\label{eq:br}
\end{equation}
where $\nu_0$ and $\gamma_0$ are the same parameters as described in equation~(\ref{eq:vvw}), and the additional terms correspond to a line coupling parameter ($\zeta_0$) and a line shift term ($\delta_0$) proportional to the gas density. The Ben-Reuven (BR) line shape is similar to the VVW line shape with the addition of a coupling element ($\zeta$) and a pressure shift term ($\delta$) \citep{10.1006/icar.2001.6622}. With both $\zeta$ and $\delta$ set to zero, the BR line shape reverts to that of VVW.

\begin{figure}[!ht]
    \centering
    \includegraphics[width=8.5cm]{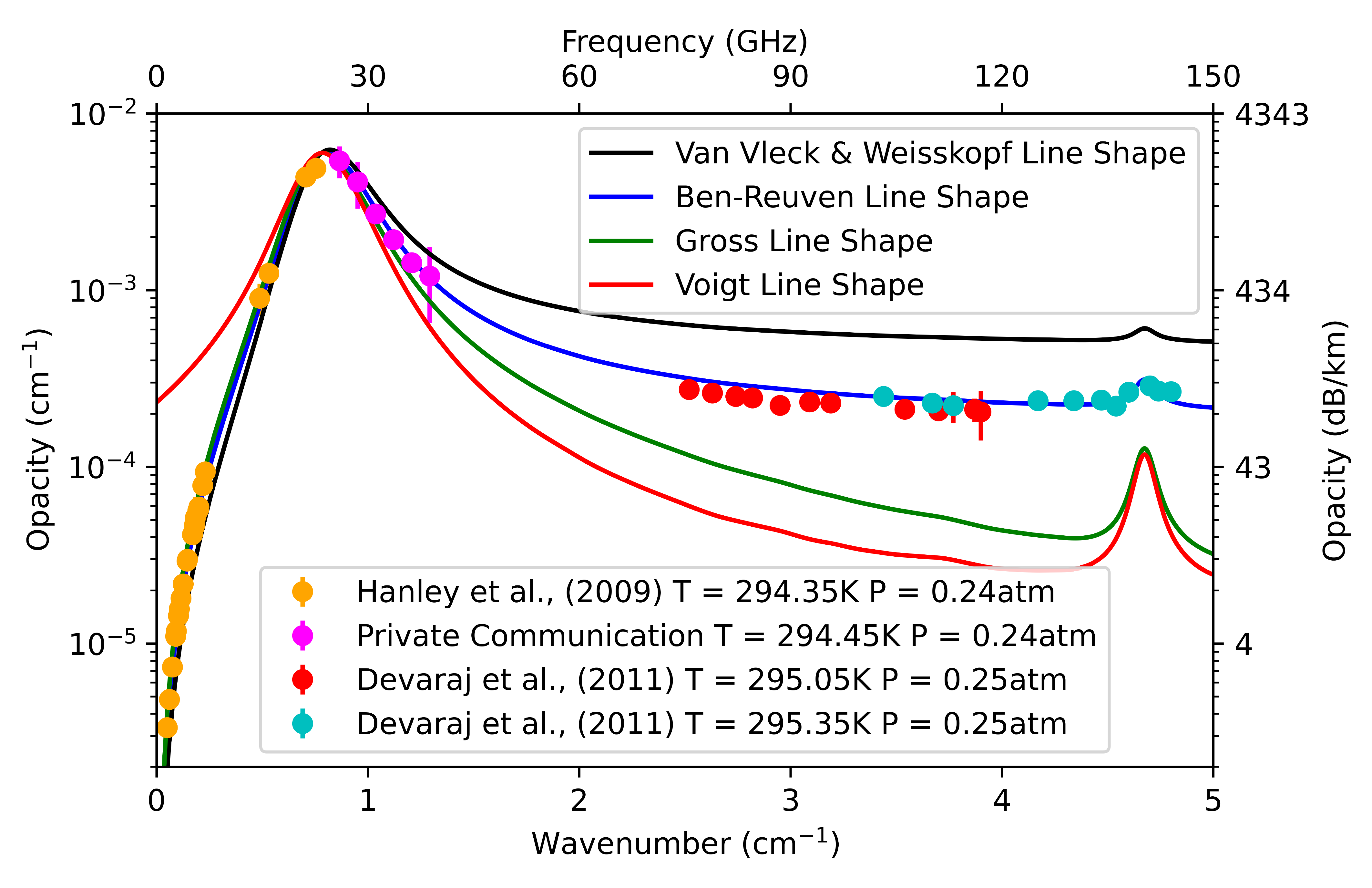}
    \caption{Opacity of pure \ch{NH3} calculated using HITRAN2020 parameters and different line shapes within HAPI at a temperature of 294.8~K and 0.25~atm vs. laboratory opacities with similar conditions. The four experimental data sets fall within $\pm$0.55~K and $\pm$0.01~atm. (\citet{10.1016/j.icarus.2010.12.010,10.1016/j.icarus.2009.02.002}, \citet{10.1016/j.icarus.2016.07.013} data and model retrieved through private communication \citep{bellotti_comm}.)}
    \label{fig:nh3_complete}
\end{figure}

When calculating spectra with HAPI, the user can choose whichever line shape they desire at any wavenumber range. A flexibility that allows users to test these differences in line shape calculations and determine the best fit choices for their particular spectrum. For low frequency transitions in most molecules, a suitable asymmetric line shape is required, such as the VVW line shape, and for rotational lines, the Gross line shape. However, the inversion mode of \ch{NH3} makes this molecule a special case, requiring the BR line shape that contains additional terms for line coupling and line shifts. An example of the line shape comparisons in the microwave region for \ch{NH3} can be seen in Fig.~\ref{fig:nh3_complete}.

The line coupling element $\zeta$ within the BR calculation is not provided in HITRAN and had to be fit to laboratory data within this work, the parameter varies by broadener type. The parameter $\zeta_{\textrm{NH}_{3}-\textrm{NH}_{3}}$ was constrained by considering only the experimental data involving pure \ch{NH3} gas, it was found to be a value of 0.35~cm$^{-1}$/atm. Similarly, $\zeta_{\textrm{NH}_{3}-\textrm{H}_{2}\textrm{O}}$ was constrained by fitting to the \ch{NH3} broadened by \ch{H2O} experimental data and was found to be 0.12~cm$^{-1}$/atm. The same process was executed for determining $\zeta_{\textrm{NH}_{3}-\textrm{H}_{2}}$, utilizing the \ch{NH3} broadened by \ch{H2} experimental data to obtain a value of 0.06~cm$^{-1}$/atm. Finally, $\zeta_{\textrm{NH}_{3}-\textrm{He}}$ was fitted to the \ch{NH3} broadened by \ch{He} laboratory data and was found to be 0.01~cm$^{-1}$/atm. 

For the rotational lines of self-broadened \ch{NH3}, a different single $\zeta$ parameter (denoted $\zeta_{\textrm{def}}$ in the codes) was implemented for every rotational transition and was fitted to laboratory data to obtain a value of 0.1~cm$^{-1}$/atm. This additional parameter was used because there is less line mixing in the rotational and roto-vibrational spectral lines due to their reduced line density when compared to the inversion transitions. The final free parameter to fit was the coefficient of the temperature dependence of the line coupling parameter ($n_{\zeta}$). A value of $n_{\zeta}$=0.55 was determined after many comparisons to experimental measurements at temperatures ranging from 200-500~K. These values are recommended while using the BR calculation for the \ch{NH3} microwave spectra.

It can be seen in Fig.~\ref{fig:nh3_complete} that a symmetric line shape (e.g., Voigt) is insufficient for these transitions, and can lead to a significant overestimate in intensity of the low frequency side of the inversion lines ($<$0.8~cm$^{-1}$), and a corresponding underestimate on the high frequency side, and for the VVW line shape the problem is reversed. Additionally, in Fig.~\ref{fig:nh3_complete} it is evident that the Gross line shape is also too low in intensity on the higher frequency side of the inversion peak. Overall, as can be seen in Fig.~\ref{fig:nh3_complete}, the BR line shape corresponds closest with available laboratory measurements and therefore is the proper line shape to use when simulating the inversion of \ch{NH3} under Jovian conditions in the microwave.

\begin{figure}[!ht]
    \centering
	\includegraphics[width=8.5cm]{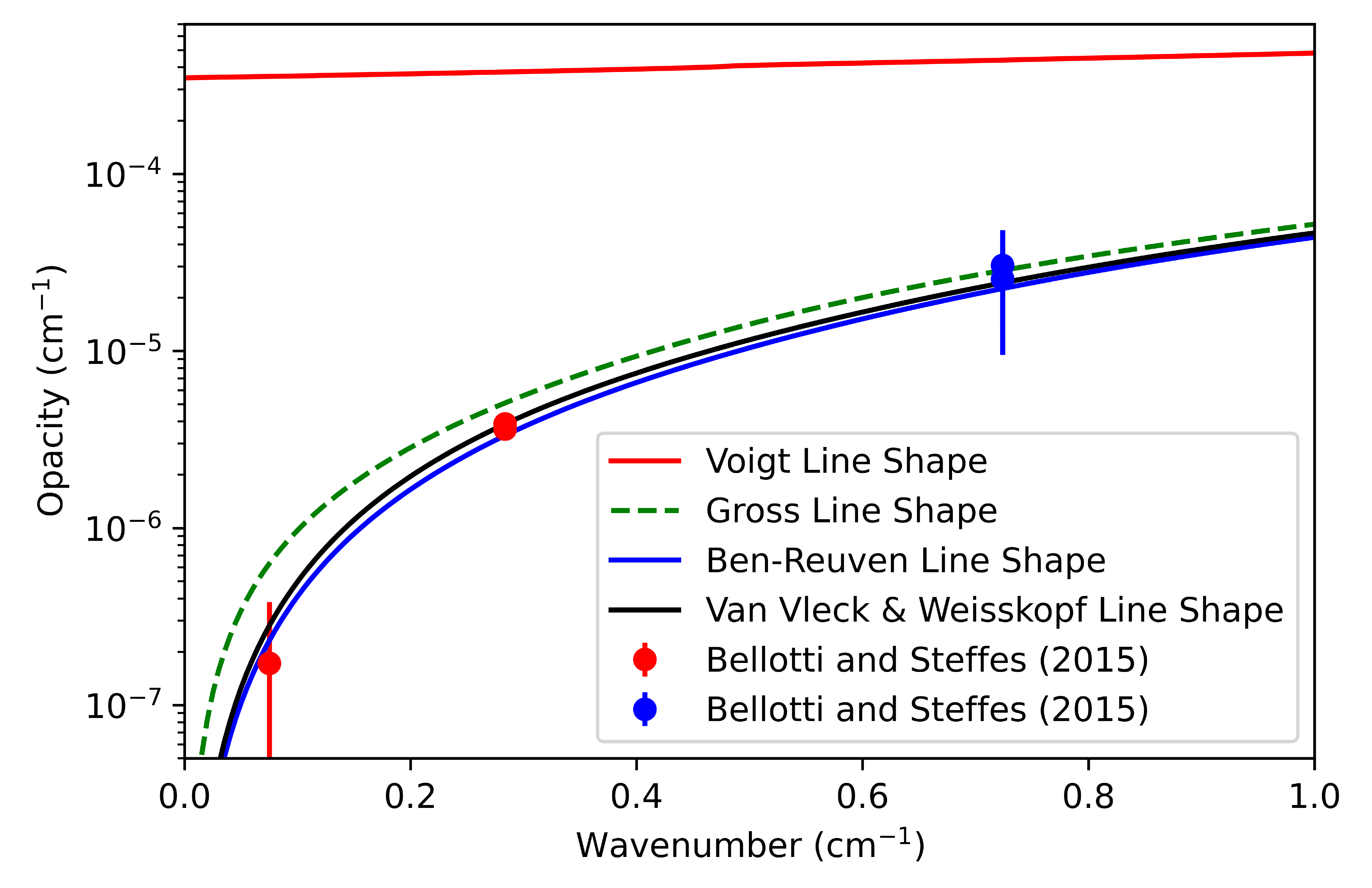}
    \caption{Opacity calculated using HITRAN2020 parameters with different line shapes within HAPI for 8.82\% \ch{SO2} and 91.18\% \ch{CO2} at a temperature at 505~K and pressure at 3.04~atm. The red data points are laboratory opacities \citep{10.1016/j.icarus.2015.03.028} with concentrations of 8.66\% \ch{SO2} broadened by 91.34\% \ch{CO2}. The blue data points are laboratory opacities \citep{10.1016/j.icarus.2015.03.028} with concentrations of 8.97\% \ch{SO2} broadened by 91.03\% \ch{CO2}. Both data sets are under the same environmental conditions of 505~K and 3.04~atm.}
    \label{fig:SO2_lineshapes}
\end{figure}

A Voigt line shape significantly overestimates low frequency opacity of \ch{SO2}, as shown in Fig.~\ref{fig:SO2_lineshapes}. The Gross line shape slightly overestimates the opacity of \ch{SO2} for the same region, however above 1~cm$^{-1}$ the asymmetric line shapes begin to converge upon one another. In Fig.~\ref{fig:SO2_lineshapes} the line shape that coincides closest with available laboratory opacities at all pressures, is the BR line shape (with the $\zeta_0$ parameter set to zero). It was found in this work that the VVW line shape applies \ch{CO2} pressure-induced line shifts poorly at higher pressure conditions. Unlike \ch{NH3} and \ch{PH3} where there are no shifts of broadening parameters included, for \ch{SO2} there are pressure-induced \ch{CO2} line shifts which must be accounted for correctly. The BR line shape includes a shift parameter within its line shape calculation ($\delta_0$) which properly applies the \ch{CO2} line shifts at higher pressure conditions. Therefore, in this work, for all \ch{SO2} broadened by \ch{CO2} calculations the BR line shape is used with $\zeta_0 = 0$. 

\begin{figure}[!ht]
    \centering
	\includegraphics[width=8.5cm]{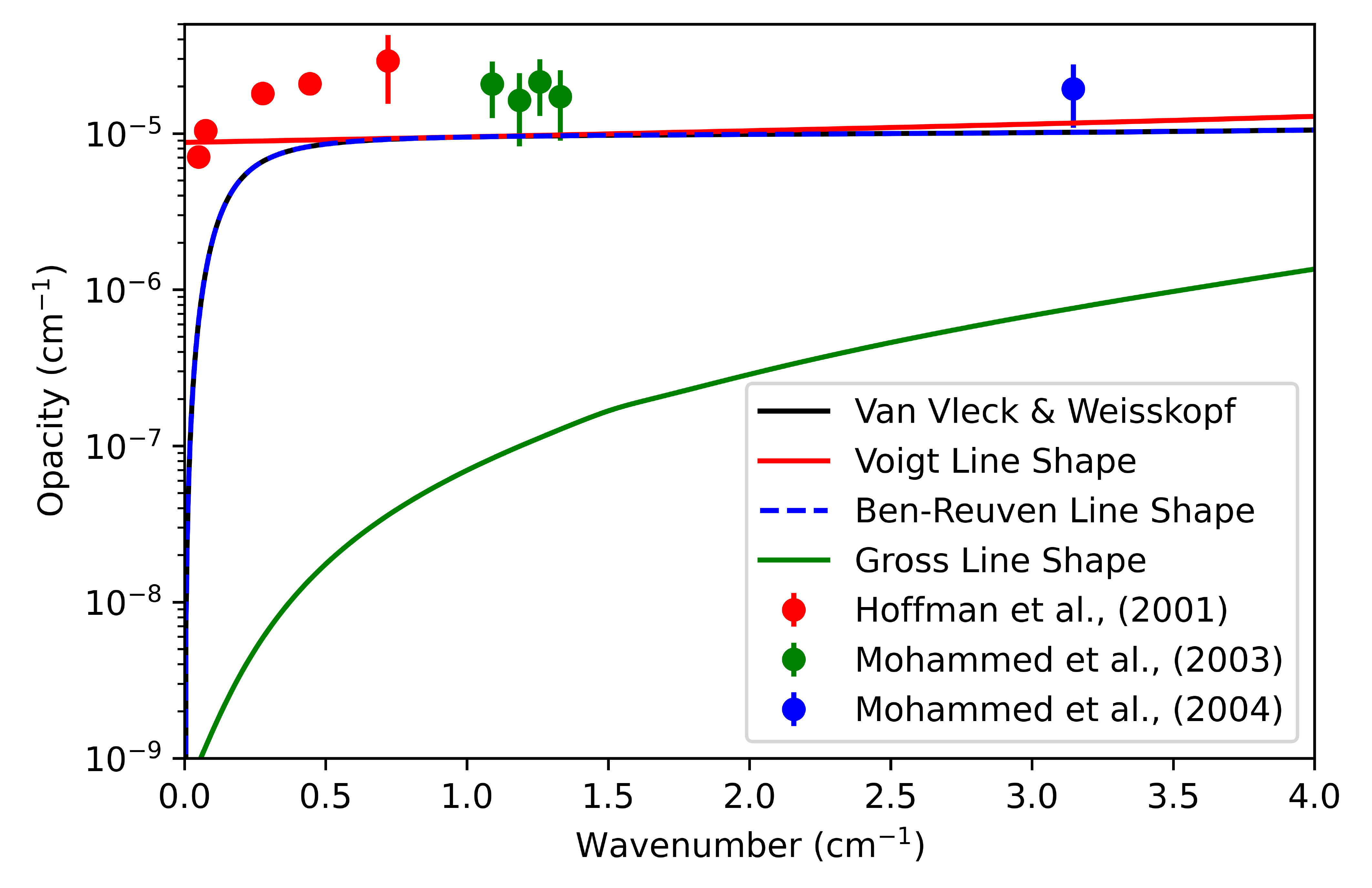}
    \caption{\ch{PH3} opacity calculated from HITRAN2020 parameters with different line shapes within HAPI compared to laboratory opacities from \citep{10.1006/icar.2001.6622,10.1016/j.icarus.2003.09.003,10.1029/2003JE002201}. Opacity is calculated at a temperature at 293~K and pressure at 1.94~atm with gas mixtures of 8.2\% \ch{PH3} broadened by 9.2\% \ch{He} and 82.6\% \ch{H2}. The gas mixtures in each data set are identical and the three experimental measurement sets fall within $\pm$1.0~K and $\pm$0.05~atm.}
    \label{fig:PH3_lineshapes}
\end{figure}

Fig.~\ref{fig:PH3_lineshapes} demonstrates how different line shape calculations simulate \ch{PH3} spectra. It is apparent in the figure that the Gross line shape, underestimates the opacity of \ch{PH3} at low frequencies. Note that the VVW line shape and the BR line shape calculations are identical and therefore appear on top of one another in the figure. This is due to the lack of fitted line coupling terms ($\zeta$) and of pressure-induced shifts ($\delta$) due to pressure of \ch{H2} and \ch{He} within the BR calculation. With these parameters set to zero the BR function reduces to VVW. In the case of \ch{NH3} the line coupling terms per broadener are fit to specific values and this causes a distinct difference between the BR and VVW line shapes. In the case of \ch{SO2} there are \ch{CO2} line shifts included within the \ch{CO2} broadening parameters for the \ch{SO2} molecule, which causes a slight difference between the BR and VVW line shapes in Fig.~\ref{fig:SO2_lineshapes}. Previous works which utilized a VVW line shape for modeling the opacity of \ch{PH3} within the microwave region include \citet{10.1029/2003JE002201,10.1016/j.icarus.2003.09.003,10.1006/icar.1999.6158,10.1006/icar.2001.6622}. In this work, the VVW line shape is chosen for \ch{PH3} calculations. 

\newpage

\section*{\small Results}\vspace{-1.5mm}\hrule\vspace{3mm}\label{sec:results}
The following comparisons to laboratory measurements are used to demonstrate the efficacy of HAPI to calculate opacities under Jovian and Venusian-like atmospheric conditions. These results include calculated spectra of \ch{NH3}, \ch{SO2} and \ch{PH3} using line parameters from HITRAN2020. Each spectrum is calculated using HAPI and the microwave line shapes (e.g., Van Vleck \& Weisskopf, Ben-Reuven) presented in this work. The calculated spectra have been compared to available experimental opacities with error bars taken from the cited text in order to validate the accuracy of the HAPI calculations. Environmental conditions and mixture concentrations vary throughout the comparisons and the spectral regions are dependent on the conditions of the experimental observations. For each comparison, the lower panel depicts the residuals from the available laboratory measurements to the HAPI calculated opacities and serve as additional quantitative evidence of HAPI's excellent performance. Additional comparisons are provided in the supplementary materials section in conjunction with publication of this work in the Monthly Notices of the Royal Astronomical Society (MNRAS).

\subsection*{\small \ch{NH3} Comparisons}\vspace{-1.5mm}\hrule\vspace{3mm}\label{sec:nh3}

\begin{figure}[!ht]
\centering
\begin{tabular}{c}
(a) \includegraphics[width=7.5cm]{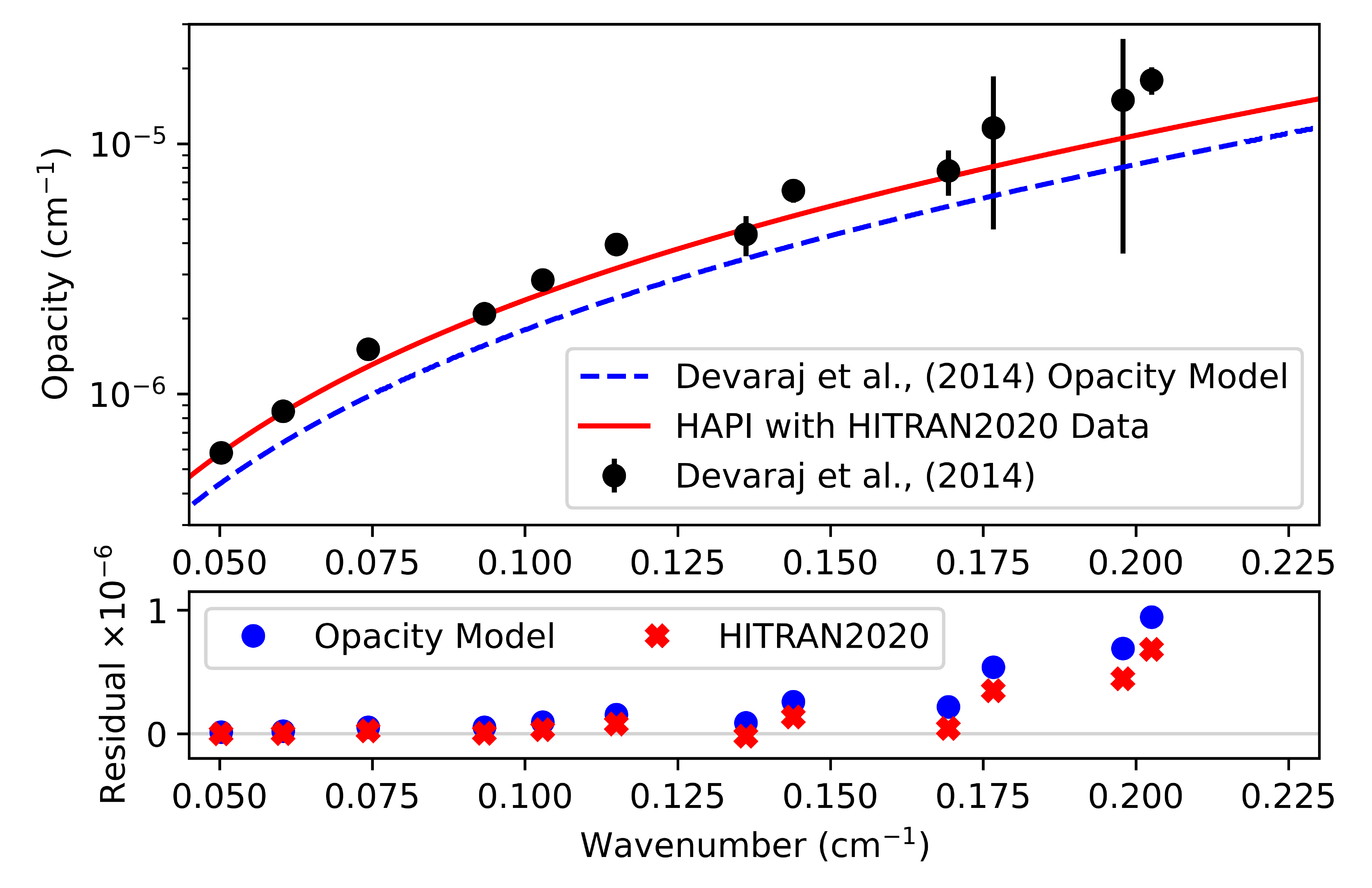}
(b) \includegraphics[width=7.5cm]{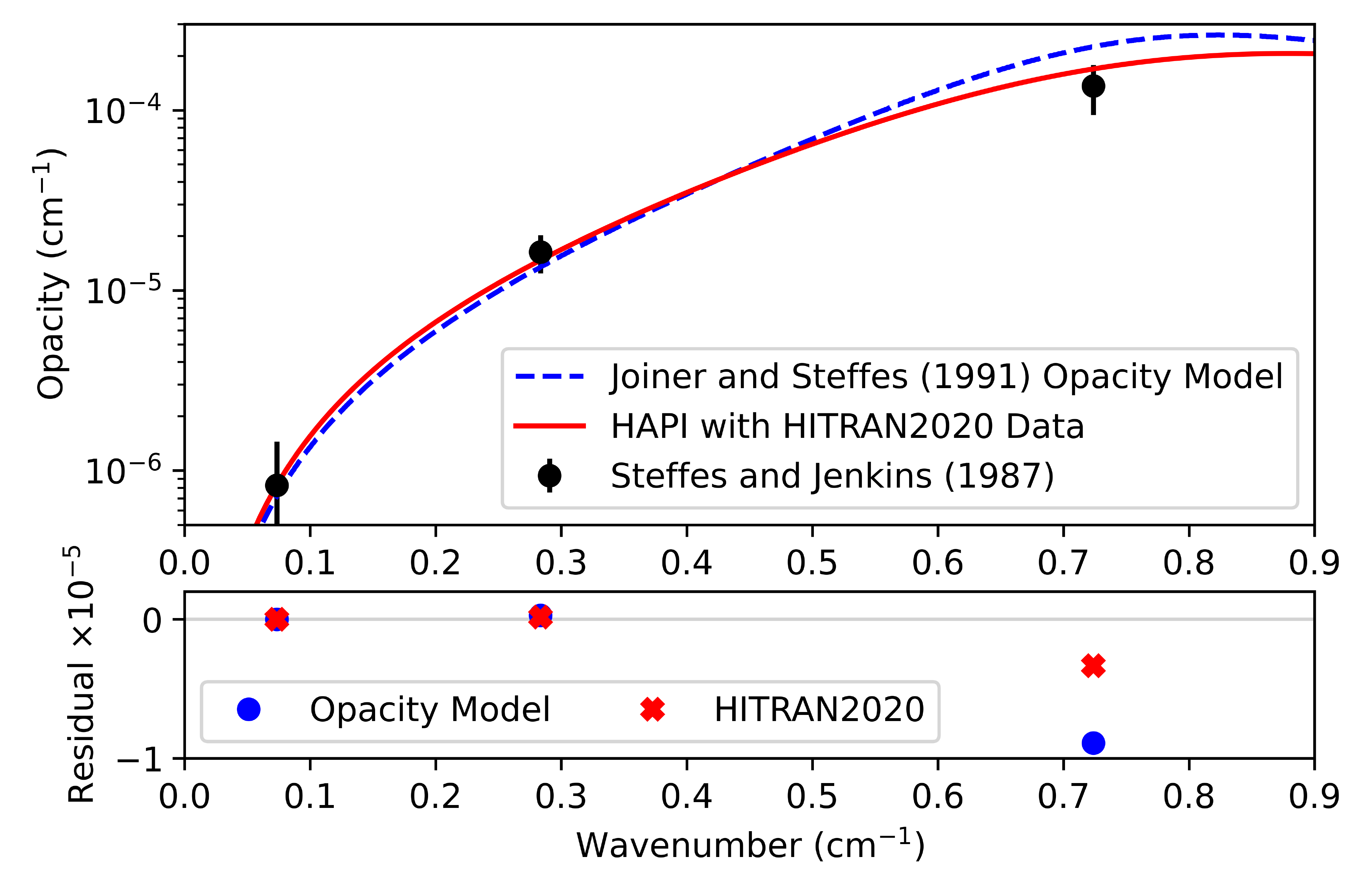} \\
(c) \includegraphics[width=7.5cm]{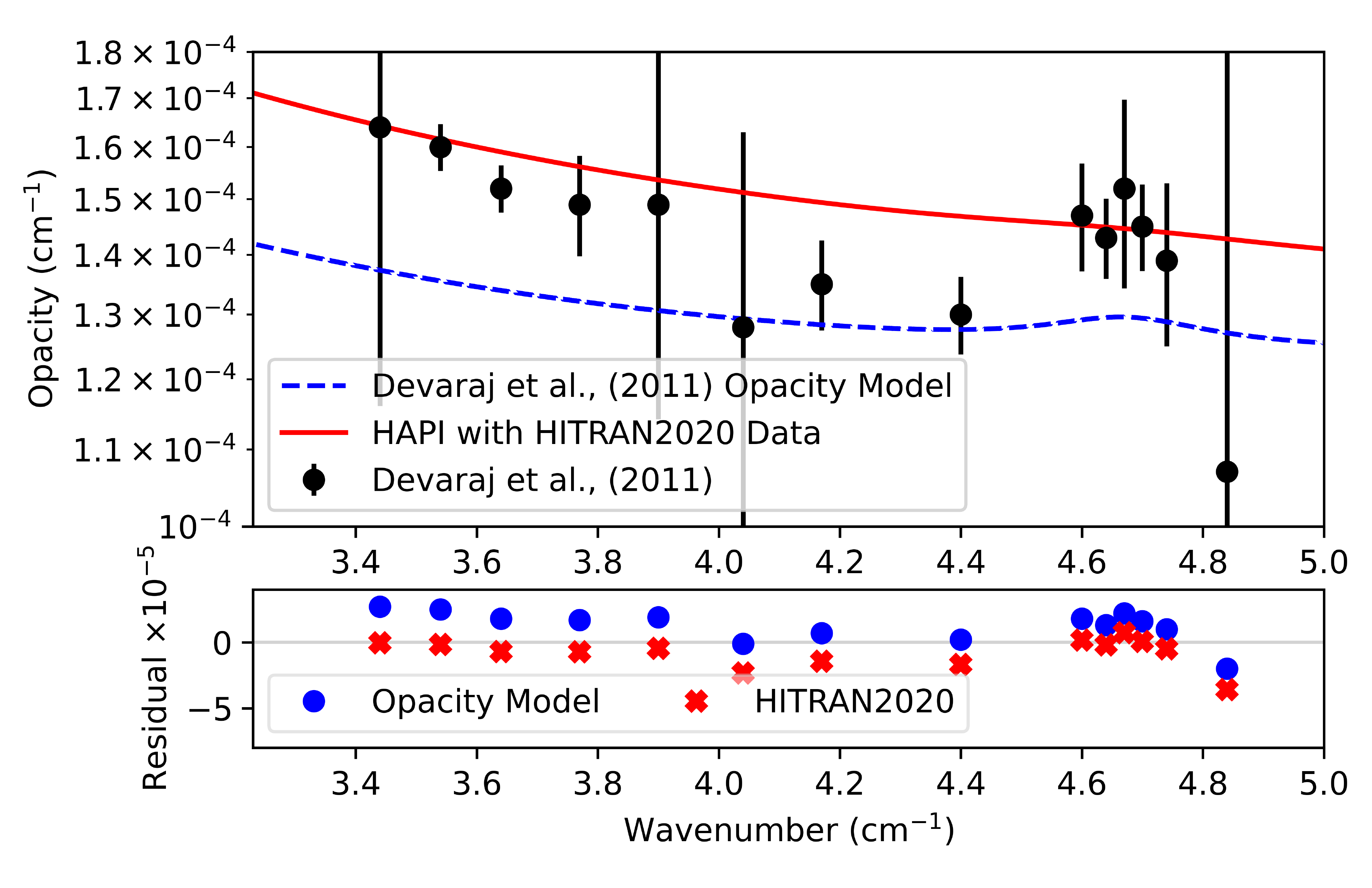}
\end{tabular}
\caption{(a) Opacity calculated using HITRAN2020 parameters and HAPI vs. experimental measurements and model \citep{10.1016/j.icarus.2014.06.017} of (17.21\%) \ch{NH3} with (82.79\%) \ch{H2O} at a temperature of 373.05~K and a pressure of 0.47~atm. (b) HAPI vs. experimental opacities and model \citep{10.1029/91JE01740,10.1016/0019-1035(87)90116-3} of (0.37\%) \ch{NH3} with (90\%) \ch{H2}, (9.63\%) \ch{He} at a temperature of 300~K and a pressure of 4.10~atm. (c) Opacity calculated using HAPI vs. experiment opacities and model \citep{10.1016/j.icarus.2010.12.010} of (1.86\%) \ch{NH3} with (84.8\%) \ch{H2} and (13.34\%) \ch{He} at a temperature of 221~K and a pressure of 2.74~atm. The lower panels indicate the residuals of the HAPI calculations and opacity models from the experimental opacities.}
\label{fig:nh3}
\end{figure}

Fig.~\ref{fig:nh3} demonstrates the effect of a variety of broadening parameters on the microwave spectrum of \ch{NH3} at a range of temperatures and pressures. Each comparison shows HAPI's \citet{10.1103/PhysRev.145.7} calculation and modeling performance at different spectral ranges compared to experimental opacities and available corresponding opacity models \citep{10.1016/j.icarus.2014.06.017,10.1016/j.icarus.2010.12.010,10.1016/0019-1035(87)90116-3,10.1029/91JE01740}. In Fig.~\ref{fig:nh3} (a) it can be seen that the laboratory measurements toward lower frequencies of the inversion peak coincide with the calculated spectrum from HAPI. The corresponding residuals demonstrate how the HAPI calculation consistently provides the best match to observation. This result can be attributed in part to the robust \ch{H2O} broadening data in HITRAN for \ch{NH3}. In Fig.~\ref{fig:nh3} (b) laboratory opacities for gas mixture concentrations of \ch{NH3} broadened by \ch{H2} and \ch{He} are shown, these are the major constituents in Jupiter's atmosphere. In Fig.~\ref{fig:nh3} (c) the spectral region at higher frequencies, beyond the primary \ch{NH3} inversion peak, is shown. The HAPI calculated opacities once again provide the smallest residuals for these conditions, suggesting that not only is HITRAN data consistent with laboratory measurements within the inversion mode but this continues up to the higher frequencies where comparisons are available (5~cm$^{-1}$). The accuracy of HAPI calculations in Fig.~\ref{fig:nh3} (b) and (c) can be attributed to accurate \ch{H2} and \ch{He} broadening values and updates provided by HITRAN2020. Further comparisons for the opacity of \ch{NH3} at a range of pressures (0.17-13.80~atm), temperatures (193-501.31~K), and mixtures (containing \ch{H2}, \ch{He}, and \ch{H2O}) are provided in the supplementary material in conjunction with publication of this work in the Monthly Notices of the Royal Astronomical Society (MNRAS).

\newpage
\subsection*{\small \ch{SO2} Comparisons}\vspace{-1.5mm}\hrule\vspace{3mm}\label{sec:so2}

\begin{figure}[!ht]
\centering
\begin{tabular}{c}
(a) \includegraphics[width=7.5cm]{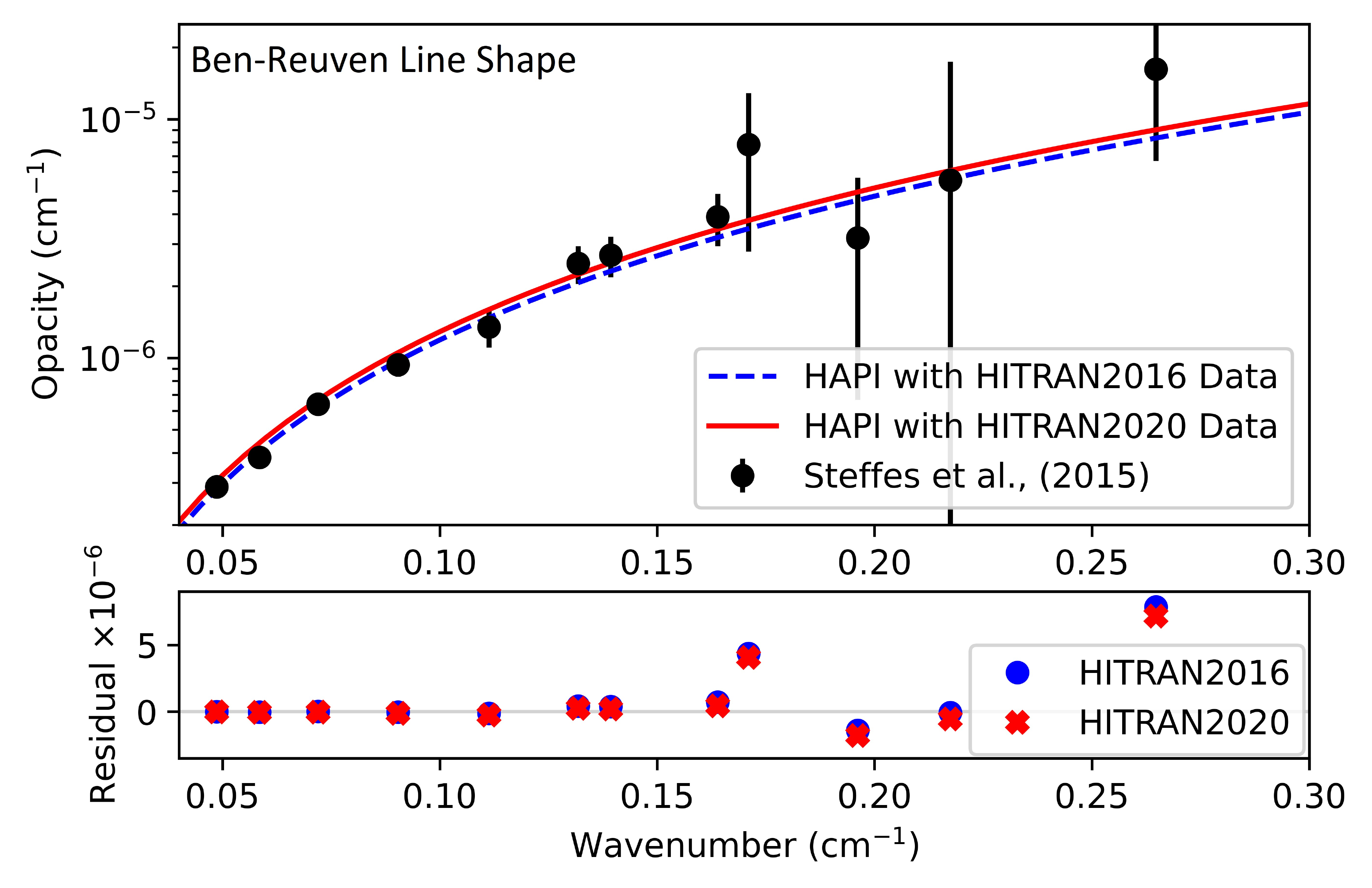}
(b) \includegraphics[width=7.5cm]{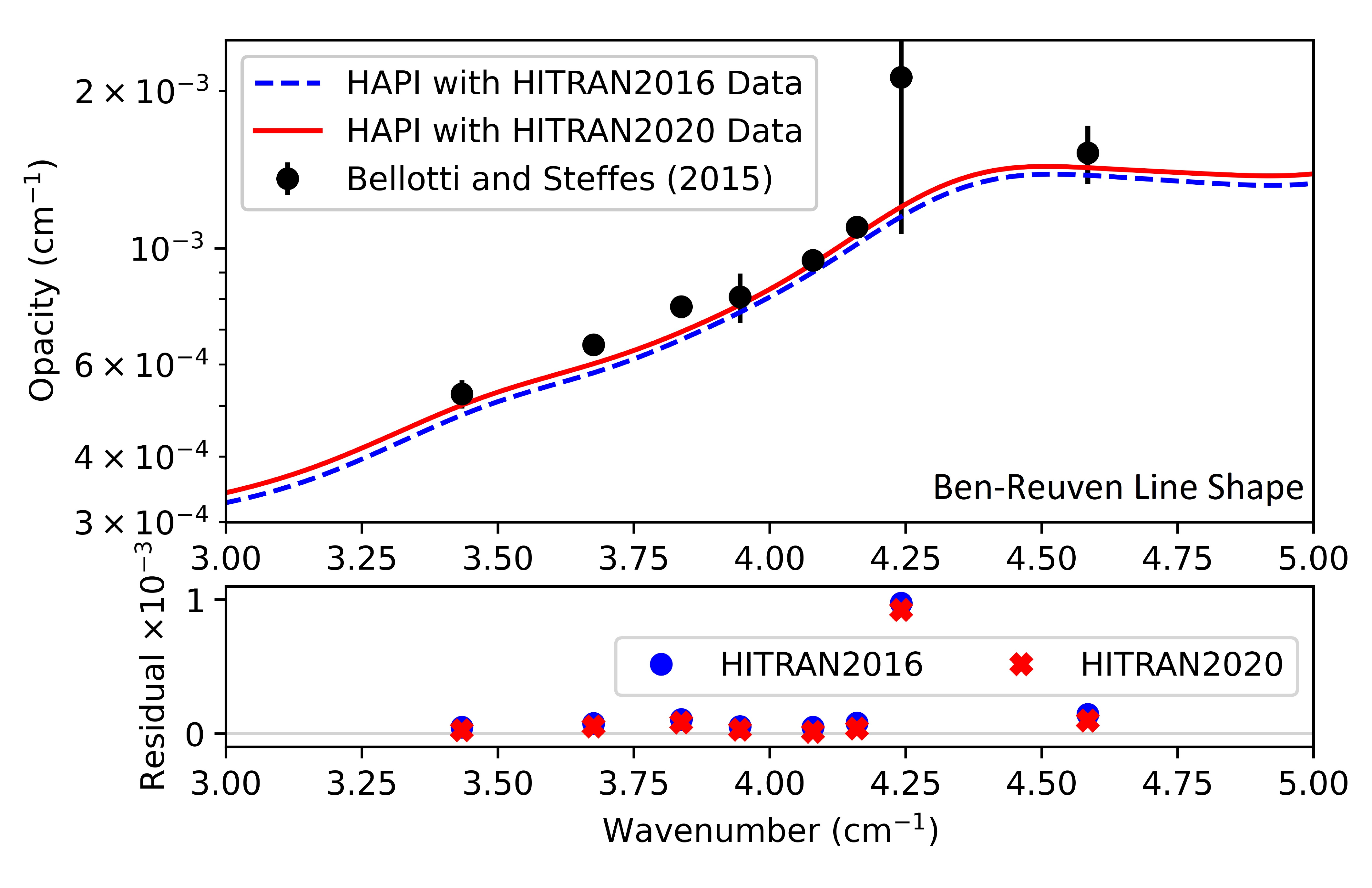} \\
(c) \includegraphics[width=7.5cm]{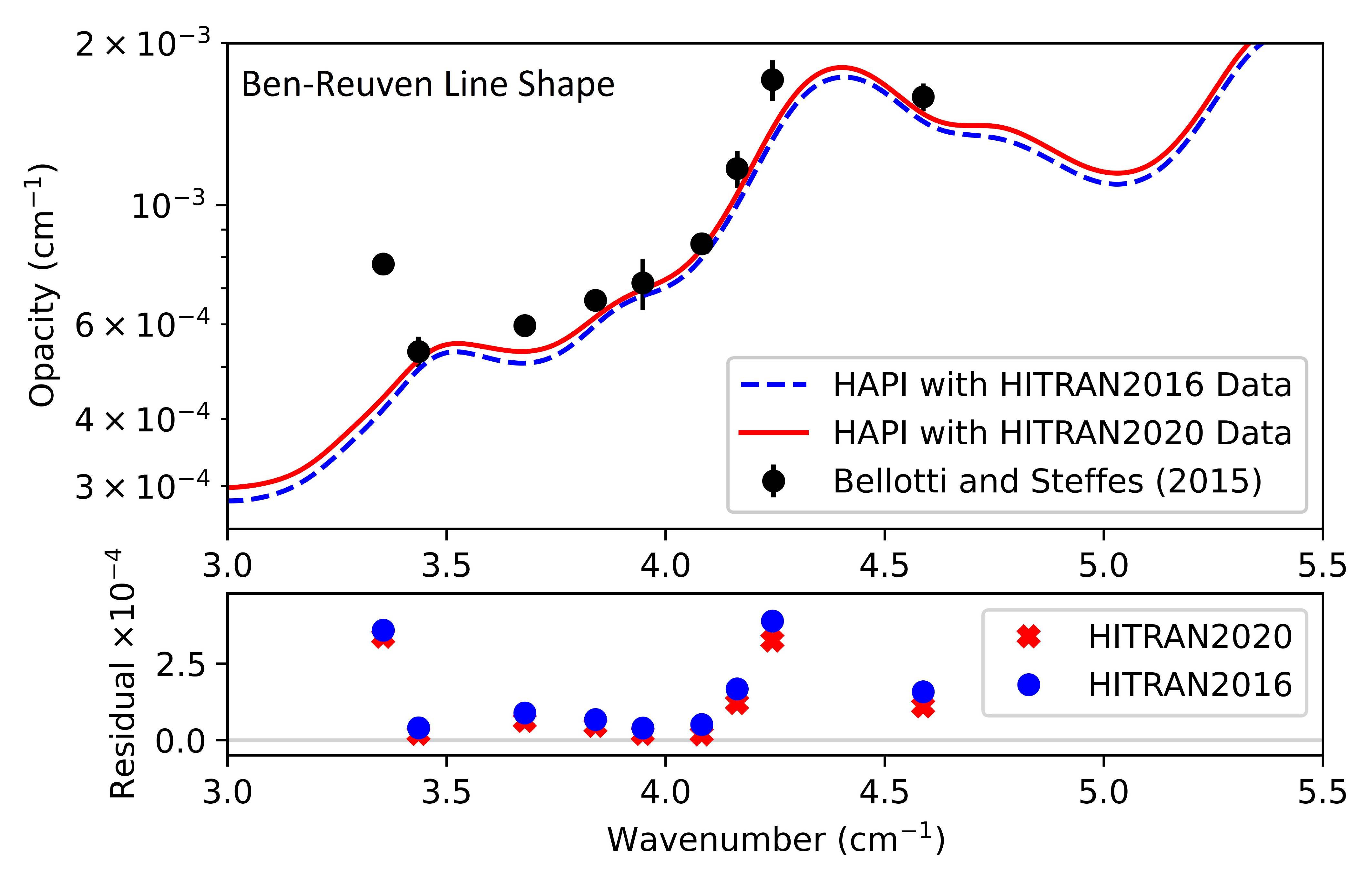}
\end{tabular}
\caption{Opacity calculated using HAPI and HITRAN2016 \& HITRAN2020 data using a Ben-Reuven line shape vs. experiment opacities \citep{10.1016/j.icarus.2015.03.028,10.1016/j.icarus.2014.09.012}. (a) Opacity calculated for (0.24\%) \ch{SO2} with (99.76\%) \ch{CO2} at a temperature of 435.27~K and a pressure of 90.79~atm. (b) HAPI vs. experiment measurements of (3.87\%) \ch{SO2} with (96.13\%) \ch{CO2} at a temperature of 343.95~K and a pressure of 1.94~atm. (c) Opacity calculated using HAPI vs. experiment opacities of (8.26\%) \ch{SO2} with (91.74\%) \ch{CO2} at a temperature of 344.65~K and a pressure of 0.91~atm. The lower panels indicate the residuals of the HAPI calculations from the experimental measurements.}
\label{fig:so2}
\end{figure}

Fig.~\ref{fig:so2} displays the use of the \ch{CO2} broadening parameters on the microwave spectrum of \ch{SO2} at various environmental conditions. Each figure includes the HAPI calculated opacity (using the BR line shape) compared to the available laboratory opacities \citep{10.1016/j.icarus.2015.03.028,10.1016/j.icarus.2014.09.012}. In these comparisons two HAPI spectra are generated using HITRAN2020 and HITRAN2016, respectively. The differences demonstrate improvements that the new \ch{CO2} broadening data have made in the microwave region. In Fig.~\ref{fig:so2} (a) the pressure and temperature conditions are very high. Therefore, excellent performance under these circumstances suggest that HITRAN data can be used to accurately calculate \ch{SO2} opacities at relatively high pressure and temperature conditions. In Fig.~\ref{fig:so2} (b) the spectral range is at higher frequencies and the HAPI calculated spectrum compares well to the laboratory opacities in this region. Fig.~\ref{fig:so2} (c) is similar to Fig.~\ref{fig:so2} (b) except it is at a lower pressure and therefore the underlying transitions are more apparent in the spectrum. As can be seen in Fig.~\ref{fig:so2} (c), at a lower pressure the improvement in the residuals between HITRAN2020 and HITRAN2016 is clearly distinct. The excellent performance of the HAPI calculations in these comparisons can be attributed to the updated \ch{CO2} broadening parameters. Further comparisons for the opacity of \ch{SO2} at a range of pressures (0.02-89.66~atm), temperatures (295-505~K), and mixtures (containing \ch{CO2} and \ch{N2}) are provided in the supplementary material in conjunction with publication of this work in the Monthly Notices of the Royal Astronomical Society (MNRAS).

\newpage

\subsection*{\small \ch{PH3} Comparisons}\vspace{-1.5mm}\hrule\vspace{3mm}\label{sec:ph3}

\begin{figure}[!ht]
\centering
\begin{tabular}{c}
(a) \includegraphics[width=7.5cm]{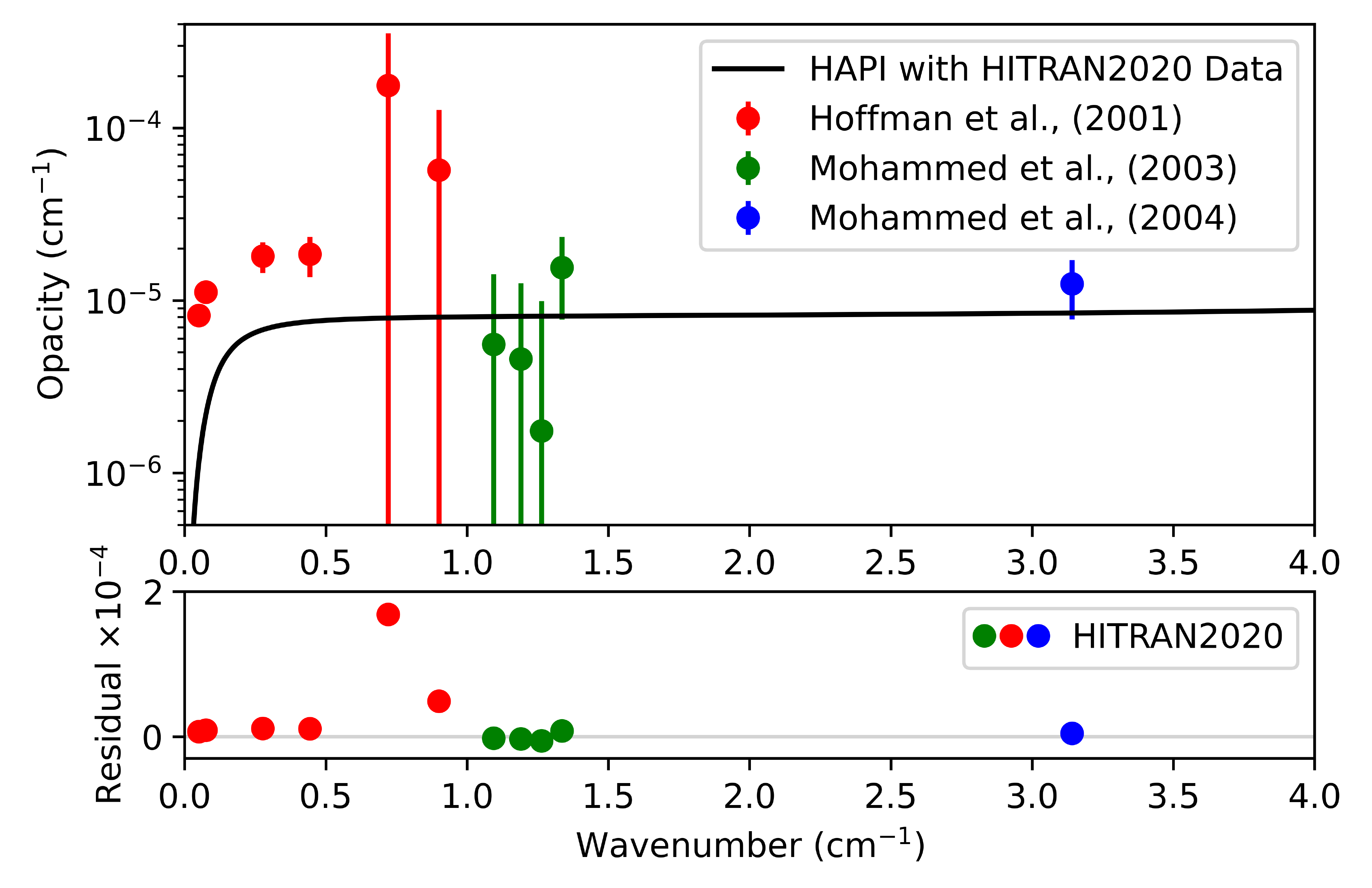}
(b) \includegraphics[width=7.5cm]{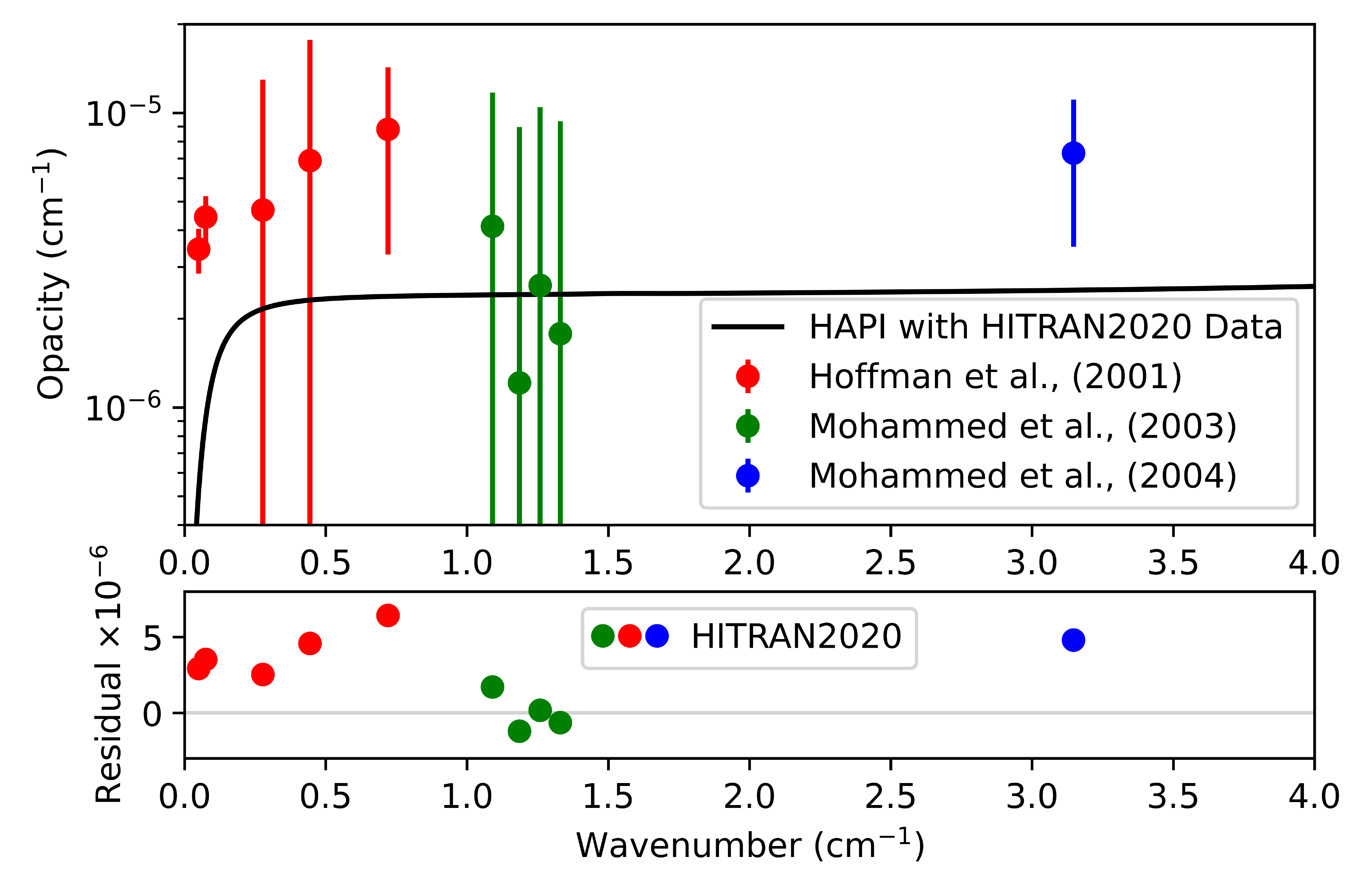} \\
(c) \includegraphics[width=7.5cm]{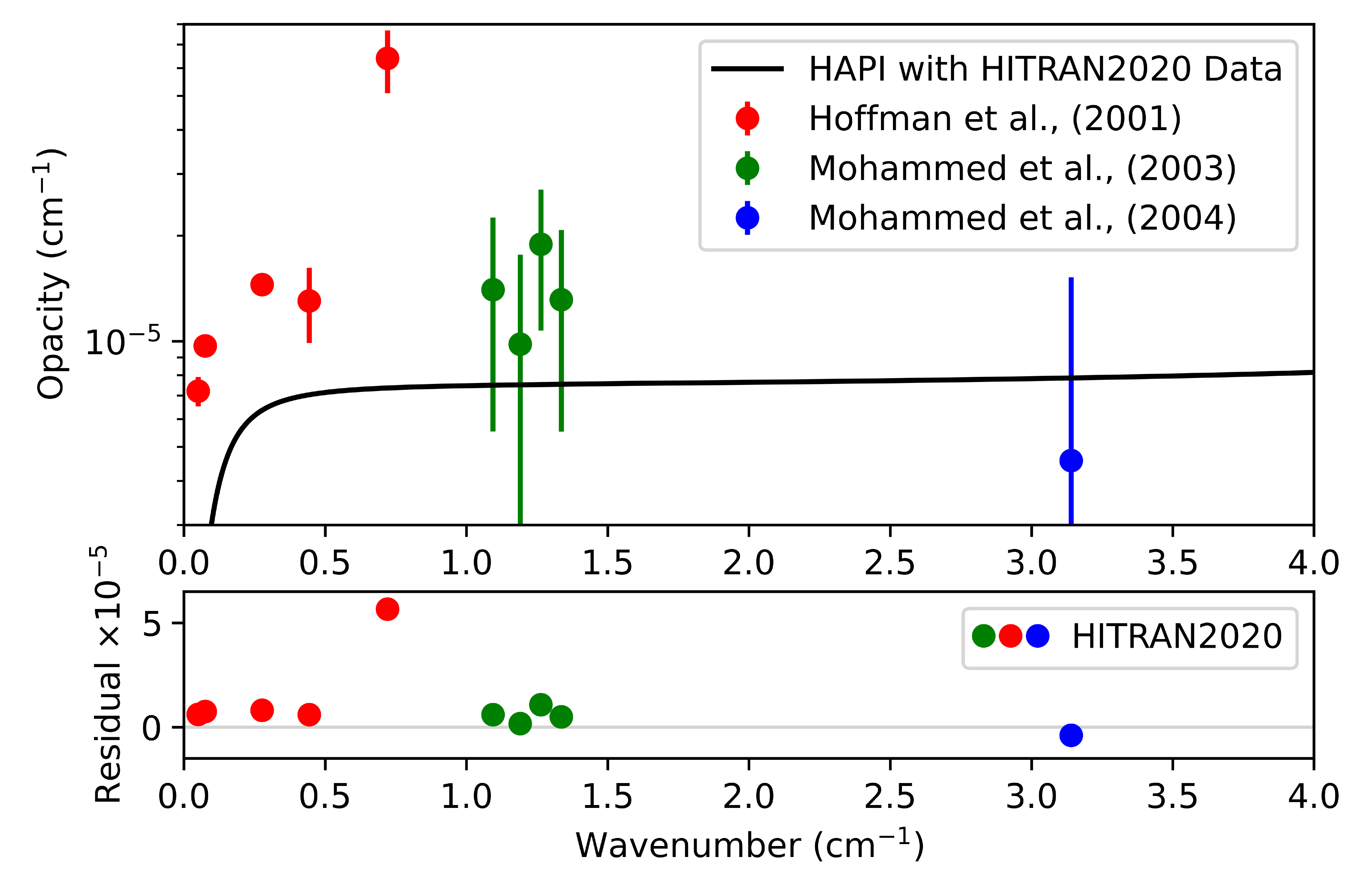}
\end{tabular}
\caption{Opacity calculated using HITRAN2020 parameters and HAPI using a Van Vleck \& Weisskopf line shape vs. experiment measurements \citep{10.1006/icar.2001.6622,10.1016/j.icarus.2003.09.003,10.1029/2003JE002201}. (a) Opacity calculated for (8.2\%) \ch{PH3} with (82.6\%) \ch{H2} and (9.2\%) \ch{He} at a temperature of 211.67~K and a pressure of 1.01~atm. (b) Opacity calculated using HAPI vs. experiment opacities of (8.2\%) \ch{PH3} with (82.6\%) \ch{H2} and (9.2\%) \ch{He} at a temperature of 295~K and a pressure of 0.976~atm. (c) HAPI vs. experiment measurements of (8.2\%) \ch{PH3} with (82.6\%) \ch{H2} and (9.2\%) \ch{He} at a temperature of 211.67~K and a pressure of 0.973~atm. The lower panels indicate the residuals of the HAPI calculations from the experimental opacities.}
\label{fig:ph3}
\end{figure}

Fig.~\ref{fig:ph3} displays the spectrum of \ch{PH3} in the region of the inversion mode. Unlike \ch{NH3}, the inversion spectrum of \ch{PH3} has never been observed despite of many attempts \citep{10.1063/1.4962259}. This is due to \ch{PH3} having a larger mass and a higher inversion energy barrier than its analogue \ch{NH3} \citep{10.1063/1.4962259} which conversely inverts easily at very low frequencies. Despite these findings \citet{10.1006/icar.2001.6622} discovered that previous attempts to approximate the opacity contribution of \ch{PH3} prior to the first 1$_{0}\leftarrow$ 0$_{0}$ line was largely underestimated. Therefore, even though the \ch{PH3} inversion spectrum is very weak, this does not negate the necessity to include the available low frequency spectroscopic data when calculating \ch{PH3} spectra in the microwave.

From the comparisons in Fig.~\ref{fig:ph3} the HAPI calculated spectra often remain within the error bars of the laboratory measurements \citep{10.1006/icar.2001.6622,10.1016/j.icarus.2003.09.003,10.1029/2003JE002201}, except at the lowest frequencies. This close correlation to laboratory opacities is in part due to the use of the VVW line shape as well as the recently added \ch{H2} and \ch{He} broadening parameters for \ch{PH3} \citep{yansplanetarybroadeners}. In Fig.~\ref{fig:ph3} (a) the gas mixtures in each data set are identical and the three laboratory experiments fall within $\pm$1.67~K and $\pm$0.07~atm. In Fig.~\ref{fig:ph3} (b) the gas mixtures in each data set are identical and the three experimental measurements fall within $\pm$3.0~K and $\pm$0.006~atm. The laboratory opacities have a more consistent trend due to the measurements from \citet{10.1006/icar.2001.6622} exhibiting less intensity variation than that of the other \ch{PH3} comparisons. This decrease in variability of the laboratory opacities allowed for the best residual between HAPI's calculation and the available laboratory data. In Fig.~\ref{fig:ph3} (c) the gas mixtures in each data set are identical and the three experimental opacities fall within $\pm$2.67~K and $\pm$0.023~atm. A further comparison for the opacity of \ch{PH3} at a pressure of 0.986~atm, temperature of 295.33~K and a mixture containing 82.6\% \ch{H2} and 9.2\% \ch{He} is provided in the supplementary material in conjunction with publication of this work in the Monthly Notices of the Royal Astronomical Society (MNRAS).

\newpage

\section*{\small Discussion}\vspace{-1.5mm}\hrule\vspace{3mm}\label{sec:discussion}

After detailed comparisons between laboratory measured opacities and HAPI calculated opacities of \ch{NH3} and \ch{SO2} in the microwave region, it was shown that the HAPI calculated opacities (using HITRAN2020 data) are able to reproduce experimental observations up to temperatures of 600~K and pressures up to 100~atm. With that, one should note that typically it is not recommended to use HITRAN data for calculations up to these limits, as many transitions that would have non-negligible contributions to the high-temperature spectra (due to thermal population of the higher-lying ro-vibrational energy levels) will be missing. Moreover, if one would want to use line-by-line data from HITRAN in the infrared part of the spectrum, 600~K may be too high to accurately reproduce the spectrum. In general, at higher temperature users are suggested to use the high-temperature molecular spectroscopic database (HITEMP) at \url{https://hitran.org/hitemp/} which contains substantially more transitions. Unfortunately, \ch{NH3}, \ch{PH3} and \ch{SO2} are not yet in HITEMP, but efforts are underway to combine validated experimental and \textit{ab initio} data (see for instance how this was done for \ch{CH4} \citep{10.3847/1538-4365/ab7a1a}) to add the high-temperature line lists for these molecules. 

\begin{figure}[!ht]
\centering
\begin{tabular}{c}
(a) \includegraphics[width=7.5cm]{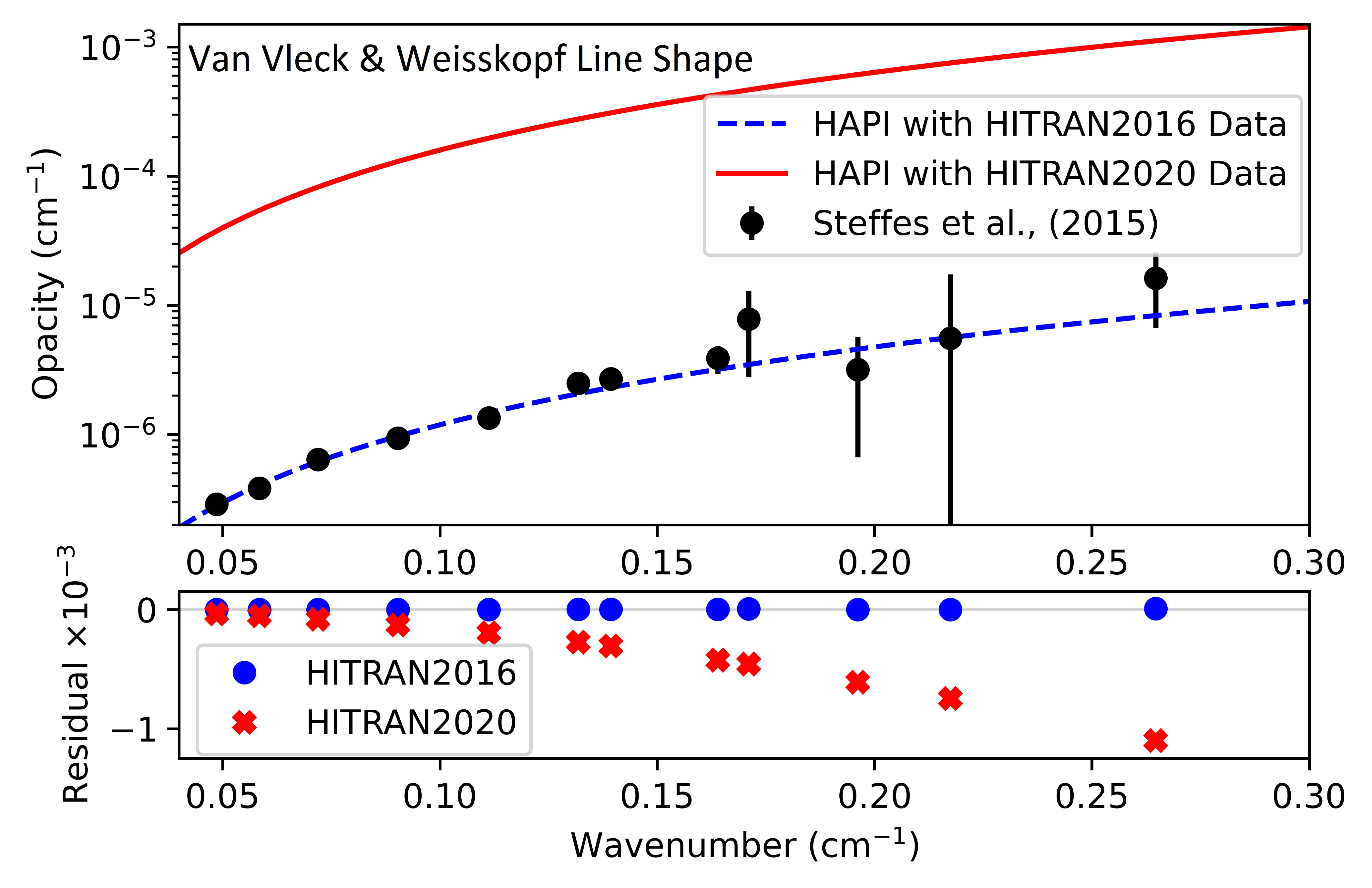}
(b) \includegraphics[width=7.5cm]{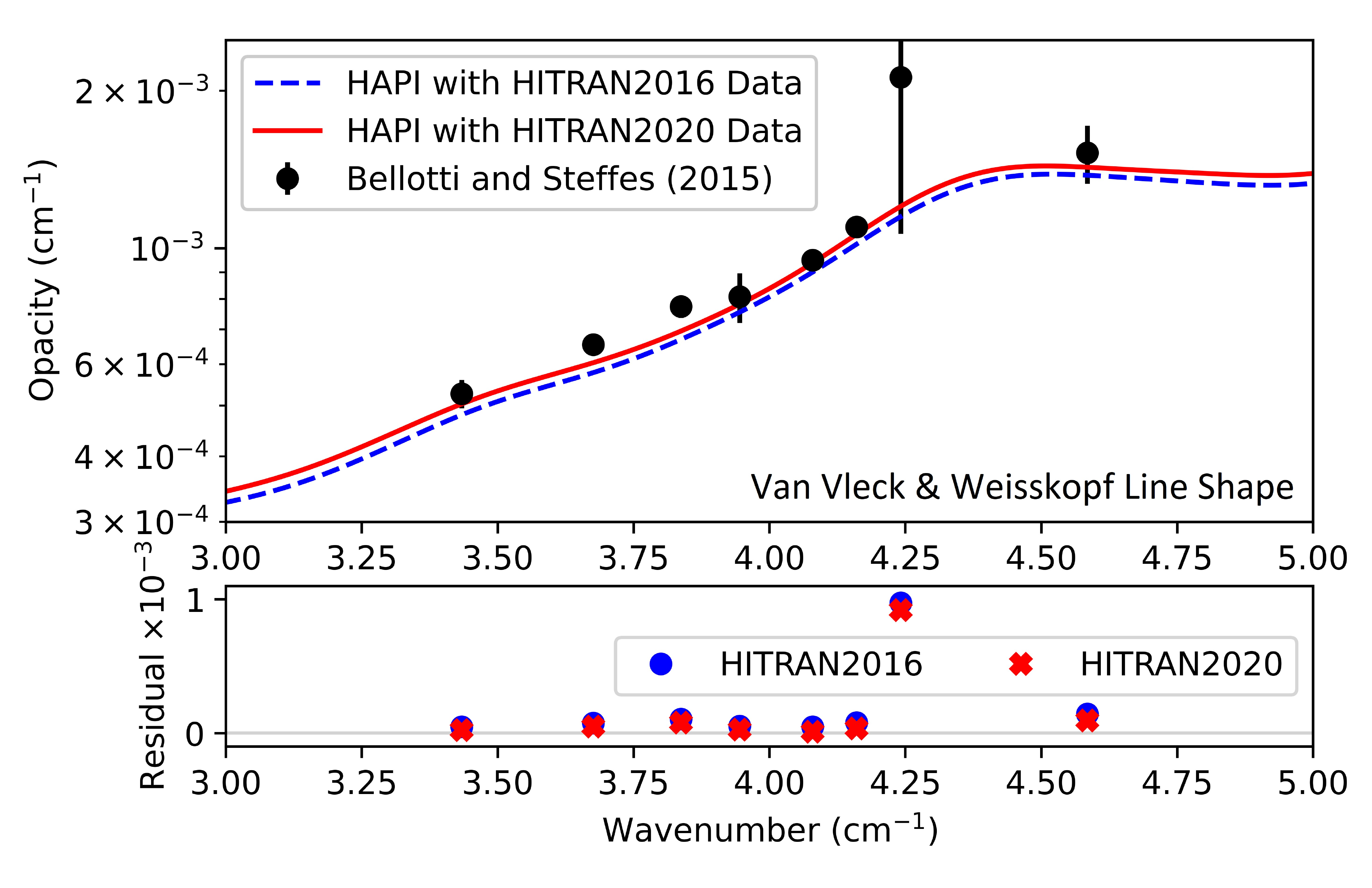} \\
(c) \includegraphics[width=7.5cm]{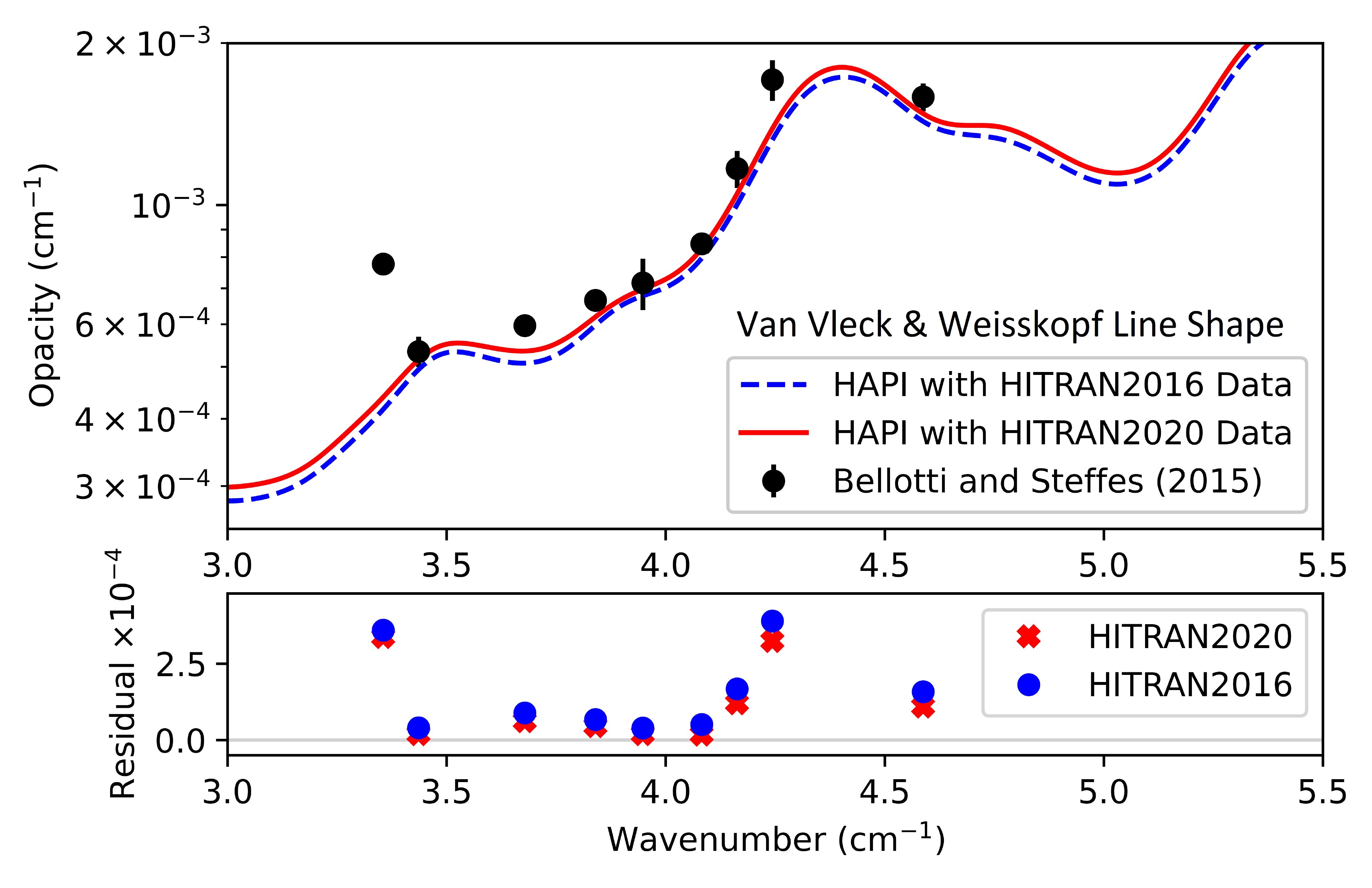}
\end{tabular}
\caption{This comparison mirrors Fig.~\ref{fig:so2} from the results section. The only difference between here and Fig.~\ref{fig:so2} is that the HITRAN2020 calculations use the Van Vleck \& Weisskopf line shape. Meanwhile in Fig.~\ref{fig:so2} the Ben-Reuven line shape was used for the HITRAN2020 calculations.}
\label{fig:so2_shifts}
\end{figure}

Upon implementation of the VVW line shape for \ch{SO2} opacity, its overall performance showed distinct issues upon comparisons with laboratory data at pressures greater than 4~atm. Although within the microwave region at pressures less than 4~atm, the VVW line shape performed well upon comparison with laboratory data. Fig.~\ref{fig:so2_shifts} parallels Fig.~\ref{fig:so2} in the results section with the primary difference being that the VVW line shape calculations are used to simulate the HITRAN2020 data in order to see their effect. Fig.~\ref{fig:so2_shifts} includes the HAPI calculated opacity (using the VVW line shape) compared to the available laboratory opacities \citep{10.1016/j.icarus.2015.03.028,10.1016/j.icarus.2014.09.012}. In these comparisons two HAPI spectra are generated using HITRAN2020 and HITRAN2016 data, respectively. Note that in HITRAN2016 the \ch{CO2} shifts were set to a value of zero for all transitions. In Fig.~\ref{fig:so2_shifts} (a) at high pressure it is evident that the inclusion of the \ch{CO2}-induced line shifts increase the calculated opacity of the spectra, which therefore causes a poor-fit to the laboratory data. In Fig.~\ref{fig:so2_shifts} (b) and (c) at lower pressures the HITRAN2020 spectra are visibly unchanged when \ch{CO2} pressure-induced line shifts are included. The VVW line shape (equation~\ref{eq:vvw}) does not explicitly include line shifts and it was applied in HAPI the same way shifts are added to the Voigt profile. This apparently fails at high pressures. The BR line shape (equation~\ref{eq:br}) incorporates the line shifts better for microwave calculations at high pressures. However, one should keep in mind that the shifts in HITRAN are not based on the fit of spectral lines with the BR line shape. Ideally they should be and the line-coupling parameter ($\zeta_0$) should be fit to high-resolution spectra when possible. In this work $\zeta_0$ is not fit due to the limited pressure data available in the literature between 4 and 30~atm in the microwave region. It may not be too dramatic of an effect as the density of rotational lines of \ch{SO2} is much less than in the inversion spectrum of \ch{NH3} and the line mixing effect would be less pronounced.

Initial comparisons to the \ch{NH3} experimental works indicated that the self-broadening of the \ch{NH3} transition at 4.67463~cm$^{-1}$ was too large in HITRAN2016. Following these comparisons, the value for self-broadening was reduced from 0.437~cm$^{-1}$/atm to 0.250~cm$^{-1}$/atm in HITRAN. The evidence for this choice is given in Fig.~\ref{fig:nh3} (c) where the strong correlation between the calculated HAPI spectrum and the laboratory opacities from \citet{10.1016/j.icarus.2010.12.010} for the 4.67~cm$^{-1}$ region is evident in both the upper and lower panels. In this work it has been demonstrated that HITRAN broadening data can be used as is to calculate the opacities of \ch{NH3}, \ch{SO2} and \ch{PH3}. 

In calculation of opacities with HAPI, this work took into account the collision induced absorption (CIA) data from HITRAN in comparisons against laboratory data. However, when plotting \ch{H2}-\ch{H2} collision complexes (originating from \citep{Abel2011,Fletcher2018}) under high pressures and temperatures, their intensity and overall influence on the microwave opacity of \ch{NH3} and \ch{PH3} were negligible. The inversion spectrum of \ch{NH3} in the microwave region is exceptionally strong in intensity and therefore dominated any inclusion of \ch{H2} collision induced absorption data. Conversely, the weak \ch{PH3} inversion spectrum was predicted to be affected by the inclusion of CIA data. However, it was determined that the inclusion of \ch{H2}-\ch{H2} collision complexes added no significant absorption to the \ch{PH3} comparisons. Similarly, for \ch{SO2} this work included \ch{CO2} collision complexes within the comparisons. However, \ch{SO2} has many dense spectral transitions within the microwave region and these effects dominate any inclusion of \ch{CO2} collision complexes at low pressure conditions. During \ch{SO2} opacity calculations, at higher pressures it was found that when CIA was included, it simulates too much opacity contribution (compared to what is seen in laboratory data). \ch{CO2}-\ch{CO2} CIA data (originating from \citet{10.1006/icar.1997.5773}) could be applied more appropriately within the \ch{SO2} calculations if the Ben-Reuven $\zeta_0$ line-coupling coefficient was fitted properly. However, in this work, there is not enough laboratory data at pressures between 4 and 30~atm within the literature, in order to properly fit the $\zeta_0$ line-coupling coefficient.

A potential issue that did not manifest itself in the comparisons of this work, is that both HITRAN and the \ch{NH3} opacity models currently assume a temperature dependence of the widths to be a single power law (equation~\ref{eq:pl}), which is known to be insufficient over a wide range of temperatures \citep{10.1016/J.JQSRT.2018.05.019}. As described in \citet{10.1016/J.JQSRT.2019.106676}, the HITRAN database has plans to adapt the double power law representation proposed by \citet{10.1016/J.JQSRT.2018.05.019}. In the present work however, a single power law was capable of sufficiently modeling the experimental observations up to 600~K. Additionally, it is likely that secondary collisions become relevant at very high pressures. However, in these test cases this effect was not pronounced enough to limit the quality of our fit to laboratory data. The pressures used in calculating opacities are taken from the experimental data and represent ``effective" ideal gas pressures calculated from \ch{H2} equation of state. Indeed, at very high pressures the compression of the gas causes departures from the ideal gas \citep{10.1016/j.icarus.2014.06.017}. 

Additional comparisons between HAPI and available \ch{NH3} and \ch{SO2} laboratory data, that are not already included in the main paper, are provided as supplementary material in conjunction with publication of this work in the Monthly Notices of the Royal Astronomical Society (MNRAS). Overall, these additional comparisons encompass more environmental conditions, gas mixture concentrations, spectral regions and they further demonstrate the excellent results attainable from HITRAN data by using HAPI. In particular, these comparisons were made possible by using new microwave line shape calculations with the HAPI program. These new calculations are in a Python programming script that is designed to be used along with HAPI \citep{10.1016/j.jqsrt.2016.03.005}. Use of these line shapes are recommended for calculations in the microwave region where the calculation includes frequencies near 0~cm$^{-1}$.

\section*{\small Conclusions}\vspace{-1.5mm}\hrule\vspace{3mm}\label{sec:conclusion}
It is well understood that the microwave emission spectrum of Jupiter's troposphere reflects the abundance and distribution of constituents such as \ch{NH3}, \ch{H2O} vapor, and aqueous \ch{NH3} clouds \citep{10.1016/j.icarus.2004.08.012}. Thus, the microwave absorption properties of \ch{NH3} and \ch{NH3} broadened by \ch{H2}, \ch{He} and \ch{H2O}, are essential for further understanding of these constituents under Jovian conditions. Many extensive laboratory studies have previously been conducted on the microwave opacity of \ch{NH3} and \ch{H2O} vapor \citep[i.e.,][]{10.1016/0019-1035(87)90116-3, 10.1016/0019-1035(89)90059-6, 10.1029/91JE01740, 10.1016/j.icarus.2003.09.003, 10.1029/2003JE002201, 10.1016/j.icarus.2009.02.002, 10.1016/j.icarus.2010.12.010, 10.1016/j.icarus.2014.06.017, 10.1016/j.icarus.2016.07.013}, which have been used for a series of opacity models. This present work utilizes the aforementioned laboratory measurements to conduct a thorough validation of HITRAN2020 data and new broadening parameters by using the HITRAN application programming interface (HAPI). This work serves to provide HITRAN users with efficient microwave line shape calculations and demonstrate HAPI's efficacy to model opacities under Jovian conditions as a test case.

The discovery of \ch{PH3} on Venus, along with \ch{PH3} identified as a potential bio-marker, has been controversial with many re-analyses (e.g. \citet{10.3847/2041-8213/abd56a,10.1051/0004-6361/202039717}) of the \citet{10.1038/s41550-020-1174-4} observations. This has created a heightened interest to study the Venusian atmosphere in the search for \ch{PH3} \citep{10.13140/RG.2.2.13819.23847}. Additionally, several re-analyses of the \ch{PH3} detection on Venus have suggested that \ch{SO2} could account for the signal \citep{10.1038/s41550-021-01422-z,10.3847/2041-8213/abde47}. Due to the fact that many of the re-analyses have utilized HITRAN \ch{SO2}-\ch{CO2} broadening data to account for their observations \citep{10.1038/s41550-021-01422-z,10.3847/2041-8213/abde47}, the potential effects of new HITRAN \ch{SO2}-\ch{CO2} broadening data have been investigated in this work. From our \ch{SO2} comparisons it is evident that the HITRAN2020 \ch{SO2}-\ch{CO2} broadening data performs better than that of HITRAN2016. On one hand this is no surprise given that the HITRAN2016 data were based off of a single measurement from 1963 \citep{10.1063/1.1733747}, on the other hand the HITRAN2016 results are still quite satisfactory and therefore unlikely to undermine conclusions of \cite{10.3847/2041-8213/abde47,10.1038/s41550-021-01422-z}. The results from the \ch{PH3} test case comparisons demonstrate that the \ch{PH3} microwave spectrum is very weak at low frequencies. This is due to \ch{PH3} having a larger mass and a higher inversion energy barrier than its analogue \ch{NH3} \citep{10.1063/1.4962259} which conversely inverts easily at very low frequencies. Additionally, for accurate analyses of \ch{PH3} detections on Venus, there is a need for measurements of \ch{PH3} broadened by \ch{CO2}, which is currently not available in the literature. 

The line shape codes utilized in this work \citep{10.1103/PhysRev.145.7,10.1103/PhysRev.97.395,10.1103/RevModPhys.17.227} are available as supplementary material in conjunction with publication of this work in the Monthly Notices of the Royal Astronomical Society (MNRAS) and will also be incorporated into the future version of HAPI, HAPI 2.0. Excellent agreement with experimental opacities in these test cases shows that using HAPI with a \citet{10.1103/PhysRev.145.7} line shape allows one to accurately calculate the opacity of \ch{NH3} under Jovian conditions and \ch{SO2} under Venusian conditions. Similarly, the opacity of \ch{PH3} can be calculated under Venusian conditions in this work by using a \citet{10.1103/RevModPhys.17.227} line shape.

This work has made hundreds of comparisons to laboratory measurements and corresponding opacity models, the majority of which have proven HAPI's efficacy at modeling \ch{NH3}, \ch{SO2} and \ch{PH3} in the microwave region. The test cases presented in this work emphasize that HAPI is available to serve as an additional tool in modeling molecular opacities under diverse planetary conditions. This is made possible by utilizing an expansive and well updated spectroscopic database, new broadening data and a sophisticated Python application to enable complicated spectra calculations for planetary atmospheres. HAPI also allows users extensive options in calculating spectra with their desired functions and outputs, which HITRAN\emph{online} also provides.

Overall, HITRAN and HAPI can be used towards modeling spectra of different molecules in atmospheres with different concentrations of dominant gases, including exoplanetary atmospheres. 

\section*{\small Acknowledgments and Funding}\vspace{-1.5mm}\hrule\vspace{3mm}
{This work was carried out in the framework of enhancing capabilities of the HITRAN database towards modeling of planetary atmospheres supported through NASA PDART grants NNX16AG51G and 80NSSC20K1059.
We acknowledge Amadeo Bellotti, who provided the most recent unified \ch{NH3} opacity model and corresponding laboratory data, referenced in this work as "Private Communication".
We respectfully acknowledge Artem Finenko who assisted with testing the inclusion of collision induced absorption data into this works comparisons.
We also wish to acknowledge Roman Kochanov, for his help with the HAPI libraries.}

\section*{\small Conflicts of Interest}\vspace{-1.5mm}\hrule\vspace{3mm}
{The authors declare that they have no known competing financial interests or personal relationships that could have appeared to influence the work reported in this paper. The funders had no role in the design of the study; in the collection, analyses, or interpretation of data; in the writing of the manuscript, or in the decision to publish the results.}

\section*{\small Bibliography\vspace{1.5mm}\hrule\vspace{3mm}}
\renewcommand{\bibsection}{}
\bibliographystyle{main.bst}
\bibliography{Main.bib}\label{bibliography}
\end{document}